\documentclass[journal,onecolumn,draftclsnofoot]{IEEEtran}
\IEEEoverridecommandlockouts
\usepackage{t1enc}
\usepackage[utf8x]{inputenc}
\usepackage[english]{babel}
\usepackage{pdfrack}
\usepackage{psfrag}
\usepackage{dsfont}
\usepackage{amsmath}
\usepackage{amssymb}
\usepackage{stmaryrd}
\usepackage[amsthm]{ntheorem-hyper}
\usepackage{flushend}

\def\iX{{\cal X}}
\def\iY{{\cal Y}}

\def\iP{{\cal P}}
\def\iS{{\cal S}}
\def\iU{{\cal U}}

\def\EE{{\mathbb E}}
\newcommand{\sumfrac}[2]{\genfrac{}{}{0pt}{}{#1}{#2}}

\theoremstyle{plain}
\newtheorem{Thm}{Theorem}
\newtheorem{all}{Proposition}
\newtheorem{Lem}{Lemma}
\theoremstyle{definition}
\newtheorem{Def}{Definition}

\theoremstyle{remark}
\newtheorem{Rem}{Remark}
\newcommand{\vX}{\mathbf X}
\newcommand{\vY}{\mathbf Y}

\newcommand{\vm}{\mathbf m}

\newcommand{\vu}{\mathbf u}
\newcommand{\vx}{\mathbf x}
\newcommand{\vy}{\mathbf y}
\newcommand{\vz}{\mathbf z}
\DeclareMathOperator{\I}{I}
\DeclareMathOperator{\HH}{H}
\DeclareMathOperator{\DD}{D}

\title{Random Access and Source-Channel Coding Error Exponents for Multiple Access Channels}
\author{L. Farkas,
        T. Kói
\thanks{This paper has been presented in part at ISIT 2013, Istanbul.}
\thanks{L\'or\'ant Farkas is with the Department
of Analysis, Budapest University of Technology and Economics,
e-mail: lfarkas@math.bme.hu}
\thanks{Tamás Kói is with the Department
of Stochastics, Budapest University of Technology and Economics and with the MTA-BME Stochastics Research Group,
e-mail: koitomi@math.bme.hu}
\thanks{The work of the authors was supported by the Hungarian National Foundation for Scientific Research, Grant OTKA K105840. The work of Tamás Kói was also supported by the project "Talent care and cultivation in the scientific
workshops of BME" (T\'AMOP - 4.2.2.B-10/1-2010-0009).}}

\begin{document}

\maketitle
\begin{abstract}
A new universal coding/decoding scheme for random access with collision detection is given in the case of two senders. The result is used to give an achievable joint source-channel coding error exponent for multiple access channels in the case of independent sources. This exponent is improved in a modified model that admits error free $0$ rate communication between the senders.   \end{abstract}

\begin{IEEEkeywords}
random access, error exponent, multiple access, source-channel coding, collision detection, $0$ rate communication
\end{IEEEkeywords}

\section{Introduction}
This paper addresses a version of the random access model of Luo, Epremides \cite{Ephremides} and Wang, Luo \cite{Jie-Luo}, which is similar to the model studied for one-way channels by Csisz\'ar \cite{Csiszar}. In the terminology of this paper, in \cite{Csiszar} the performance of a codebook library consisting of several constant composition codebooks with pre-determined rates has been analyzed. It has been shown that simultaneously for each codebook the same error exponent can be achieved as the random coding exponent of this codebook alone. This theorem is used in \cite{Csiszar} to give an achievable error exponent for joint source-channel coding (JSCC).

Here, the mentioned results of \cite{Csiszar} are generalized to (discrete memoryless) multiple access channels (MACs). A two-senders random access model is introduced, in which the senders have codebook libraries with constant composition codebooks for multiple rate choices. The error exponent of Liu and Hughes \cite{Hughes} for an individual codebook pair is shown to be simultaneously achievable for each codebook pair in the codebook libraries, supplemented with collision detection in the sense of \cite{Ephremides,Jie-Luo}. This is achieved via a universal decoder, hence, a positive answer is given to the question in \cite{Ephremides} whether or not the results there are still valid if the receiver does not know the channel. Moreover, achievable JSCC error exponents for transmitting independent sources over a MAC are given, admitting improvements when special error free $0$ rate communication is allowed between the two senders. The most direct extension of the JSCC result in \cite{Csiszar} is obtained in the latter case.

Nazari, Anastasopoulos, and Pradhan in \cite{Nazari} derive achievable error exponents for MAC's using $\alpha$-decoding rules introduced for one-way channels by Csisz\'ar and K\"orner in \cite{alpha}. In the present paper a particular $\alpha$-decoder is used, modified similarly as in \cite{Csiszar} and supplemented by a threshold criterion in order to allow collision detection. As the proofs follow \cite{Nazari} closely, it can be seen that other $\alpha$-decoders could also be appropriate, depending on actual assumptions on the analyzed models.

Note that another multiterminal generalization of the JSCC result in \cite{Csiszar} appears in Zhong, Alajaji, Campbell \cite{Alajaji}. We also mention that this paper, as \cite{Csiszar}, has connections with the topic of unequal protection of messages, see for example Borade, Nakiboglu, Zheng \cite{Borade}.

Finally we list here some connected recent papers. Luo in \cite{Luo} generalizes the earlier results in \cite{Ephremides,Jie-Luo}. Papers \cite{dispersioncikk} and \cite{yanina} by Wang, Ingber, Kochman and Shkel, Tan, Draper, respectively, analyze the models and the results of \cite{Csiszar} in the finite blocklength regime. Campo, Vazquez-Vilar, F\`abregas, Koch, Martinez in \cite{campo} analyze the JSCC exponent achieved in \cite{Csiszar} for one-way channels in greater detail.

\section{Notation} \label{notation}
The notation follows \cite{Csiszar} and \cite{Nazari} whenever possible. All alphabets are finite and $\log$ denotes logarithm to the base $2$. The set $\{1,2,\dots,M\}$ is denoted by $[M]$. In assertions involving a sequence approaching $0$, the parameters on which this sequence depends will be indicated in parantheses. An explicit form of this dependence can be extracted from the corresponding proof.

Random variables $U$, $X$, $Y$, etc., with alphabets $\mathcal{U}$, $\mathcal{X}$, $\mathcal{Y}$, etc., will be assigned several different (joint) distributions. These will be denoted by $P_{U}$, $P_{UX}$, $P_{UXY}$, etc. or $V_{U}$, $V_{UX}$, $V_{UXY}$, etc., and corresponding conditional distributions by $P_{X|U}$ or $V_{X|U}$, etc. The first notation will typically refer to a distinguished (joint) distribution, the second one refers to distributions introduced for technical purposes such as representing joint types. The family of all distributions on $\mathcal{U} \times \mathcal{X}$, say, is denoted by $\mathcal{P} (\mathcal{U} \times \mathcal{X})$, and the family of all conditional distributions on $\mathcal{X}$ conditioned on $\mathcal{U}$  by $\mathcal{P} (\mathcal{X} | \mathcal{U})$. If a multivariate distribution, say $V_{UXY} \in \mathcal{P} (\mathcal{U} \times \mathcal{X} \times \mathcal{Y})$ is given then $V_{U}$, $V_{UX}$, $V_{X|U}$, $V_{Y|U}$, etc. will denote the associated marginal respectively conditional distributions.

The type of an $n$-length sequence $\vx=x_1 x_2 \dots x_n \in \mathcal{X}^n$ is the distribution $P_{\vx} \in \mathcal{P} (\mathcal{X})$ where $P_{\vx}(x)$ is the relative frequency of the symbol $x$ in $\vx$. The joint type of two or more $n$-length sequences is defined similarly and, for $(\vu,\vx) \in \mathcal{U}^n \times \mathcal{X}^n$, say, it is denoted by $P_{(\vu,\vx)}$. The family of all possible types of sequences $\vx \in \mathcal{X}^n$ is denoted by $\mathcal{P}^n (\mathcal{X})$, and for $P \in \mathcal{P}^n (\mathcal{X})$  the set of all $\vx \in \mathcal{X}^n$ of type $P_{\vx}=P$ is denoted by $T^n_{P}$. For $\vu \in \mathcal{U}^n$ of type $P_{\vu}$, the set of conditional distributions $V_{X|U} \in \mathcal{P}(\mathcal{X}|\mathcal{U})$ for which $P_{\vu}V_{X|U} \in \mathcal{P}^n (\mathcal{U} \times \mathcal{X})$ is denoted by $\mathcal{P}^n (\mathcal{X} | P_{\vu})$. The set of all $\vx \in \mathcal{X}^n$ with $P_{(\vu,\vx)}=P_{\vu} V_{X|U}$, non-empty if and only $V_{X|U} \in \mathcal{P}^n (\mathcal{X}|P_{\vu})$, is denoted by $T^n_{V_{X|U}} (\vu)$.

Denote $\HH_V(X,Y)$, $\HH_V(U,X,Y)$, $\I_V(X \wedge Y)$ etc. the entropy and mutual information when the random variables $U$, $X$, $Y$ have joint distribution $V_{XY}$, $V_{UXY}$ etc. Denote $\I(\vx \wedge \vy)$, $\HH(\vx,\vy)$ etc. the information quantities $\I_V(X \wedge Y)$, $\HH_V(X,Y)$ etc. with $V_{XY}=P_{(\vx,\vy)}$. Moreover, we define multi-information as in \cite{Hughes} (See also \cite{Csiszar2} exercise 3.9., and \cite{watanabe}):
\begin{equation}
\I(X_1 \wedge X_2 \wedge \dots \wedge X_N|Y) \triangleq \HH(X_1|Y)+\HH(X_2|Y)+ \dots +\HH(X_N|Y)-\HH(X_1,X_2,\dots,X_N|Y).  \label{ujjeloles}
\end{equation}
Note that a similar notation is used with a different meaning in Yeung's book \cite{Yeung}.

Given a MAC $W: \mathcal{X} \times \mathcal{Y} \rightarrow \mathcal{Z}$, the pentagon
\begin{equation} \label{pentagondef}
\begin{array}{l}

\left\{
\begin{array}{l}

(R_1,R_2): 0 \leq R_1 \leq \I(X\wedge Z|U,Y), \\
0 \leq R_2 \leq \I(Y\wedge Z|U,X), R_1+R_2 \leq \I(X,Y\wedge Z|U)
\end{array}
\right\}

\end{array}
\end{equation}where $U,X,Y,Z$ have joint distribution equal to $P_{U}P_{X|U}P_{Y|U}W$, is denoted by $C[W,P_{U},P_{X|U},P_{Y|U}]$. The union of these pentagons, i.e., the capacity region of the MAC $W$, is denoted by $C(W)$.

The following elementary facts will be used (see, e.g., \cite{Csiszar2}):

\begin{align} \label{basicfact1}
  &|\iP^n(\iX)|\leq (n+1)^{|\iX|} ,\, |\iP^n(\iX|P_{\vu})| \leq (n+1)^{|\iX||\iU|}, \\
  &\frac{2^{n\HH(P)}}{(n+1)^{|\iX|}} \le |T^n_{P}|\le 2^{n\HH(P)} \text{ and } Q^n(T^n_{P}) \leq 2^{-n \DD(P \| Q)} \textnormal{ if }\,  P \in \iP^n(\iX),  \label{basicfact2}\\
  &\frac{2^{n\HH_{V}(X|U)}}{(n+1)^{|\iX||\iY|}}\leq|T^n_{V_{X|U}}(\vu)|\leq 2^{n\HH_{V}(X|U)} \, \label{basicfact3} \textnormal{ if }\,  V_{X|U} \in \iP^n(\iX|P_{\vu}), V_{UX}=P_{\vu}V_{X|U} , \\
	&W^n(\vy|\vx)=2^{-n\left(\DD(V_{Y|X}\|W|P_\vx)+\HH_{V}(Y|X)\right)} \label{basicfact4}
	\textnormal{ where $V_{XY}=P_{(\vx,\vy)}$.}
\end{align}
Here the conditional information divergence is defined by
\begin{equation} \label{Kullback}
\DD(P_{Y|X}\|W|P_{X}) \triangleq \sum_{x \in \mathcal{X}} P_{X}(x) \DD(P_{Y|X=x}\|W(\cdot|x)),
\end{equation}
where $\DD(P_{Y|X=x}\|W(\cdot|x))$ is the I-divergence between the given conditional distributions on alphabet $\mathcal{Y}$.

Finally the variational distance between distributions $P$, $Q$ on alphabet $\mathcal{X}$ is
\begin{equation} \label{TV}
||P-Q|| \triangleq \sum_{x \in \mathcal{X}} |P(x)-Q(x)|.
\end{equation}

\section{Random Access with collision detection}\label{Modell}
In this model two transmitters try to communicate over a MAC $W$ with one common receiver. The channel $W$ is unknown to the senders and may also be unknown to the receiver (but see Remarks \ref{nonuniversal2} and \ref{nonuniversal}). Both senders have multiple codebooks of block length $n$. We assume that a common auxiliary sequence $\vu$ is given, and the codewords' conditional type on $\vu$ is fixed within codebooks, but can vary from codebook to codebook.
\newpage
\begin{Def} \label{constantcomposition}
Let a finite set $\mathcal{U}$, a sequence $\vu \in \mathcal{U}^n$ of type $P_{U} \in \iP^n (\mathcal{U})$, positive integers $M_1$ and $M_2$, conditional distributions $\{ P^{i}_{X|U} \in \iP^n (\iX|P_{U}), {i}\in [M_1] \}$, $\{ P^{j}_{Y|U} \in \iP^n (\iY|P_{U}), {j}\in [M_2] \}$, rates $\{ R_1^{i} , i \in [M_1]\}$ and  $\{R_2^{j}, {j}\in [M_2] \}$ be given parameters.

A constant composition codebook library pair of length $n$ with the above parameters is a pair $(\mathcal{A},\mathcal{B})$ where $\mathcal{A}$ and $\mathcal{B}$  consist of constant composition codebooks $(A^{1},A^{2},\dots,A^{M_1})$ resp. $(B^{1},B^{2},\dots,B^{M_2})$ such that $A^{i}=\{\vx^{i}_1,\vx^{i}_2, \dots \vx^{i}_{N_1^i}  \} $ and $B^j=\{\vy^{j}_1,\vy^{j}_2, \dots \vy^{j}_{N_2^j}  \}$ with $\vx^{i}_{a} \in T^{n}_{P_{X|U}^{i}} (\vu)$ and $\vy^{j}_{b} \in $ $T^{n}_{P_{Y|U}^{j}} (\vu)$, ${i} \in [M_1]$, $j \in [M_2]$, $N_1^i = \left\lfloor 2^{nR_1^i} \right\rfloor$, $N_2^j = \left\lfloor 2^{nR_2^j} \right\rfloor$,  $a \in [N_1^i]
$, $b \in [N_2^j]$.
\end{Def}
\begin{Rem}
In the above definition all parameters can depend on $n$. Note that $\mathcal{U}$ is fixed in Theorem \ref{collision} and it is assumed that $M_1$ and $M_2$ grow at most subexponentially with $n$.
\end{Rem}

Before sending messages, each transmitter chooses one of its codebooks independently of the other sender. Denote this selection by $(i,j) \in [M_1]\times [M_2]$. The transmitters do not share the result of their selections with each other, neither with the receiver. The senders send codewords $\vx_{a}^{i}$, $\vx_{b}^{j}$. The decoder's output $\hat\vm$ is either a quadruple $(\hat i,\hat a,\hat j,\hat b)$ or "collision". The receiver is required to decode quadruple $(i,a,j,b)$ if the rate pair $(R_1^{i},R_2^{j})$ of the chosen codebooks is in the interior\footnote{Here, interior is meant in the relative topology of the nonnegative quadrant. In particular, the rate pairs $(R_1, 0)$ with $0 \le R_1 < I(X\wedge Z|U,Y)$ belong to the interior of the pentagon (\ref{pentagondef}).} of $C[W,P_{U},P^{i}_{X|U},P^{j}_{Y|U}]$ and to declare "collision" otherwise; cf. \cite{Ephremides}. Hence, two types of error are defined, one of them should be small according as $(R_1^{i},R_2^{j})$ is in the interior of $C[W,P_{U},P^{i}_{X|U},P^{j}_{Y|U}]$ or not.
\begin{Def}
For the codebooks $(A^i, B^j)$, the average decoding error probability $Err_d (i,j)$ is
\begin{equation}
\frac{1}{N_1^{i}N_2^j}\sum_{a=1}^{|A^i|}\sum_{b=1}^{|B^j|} Pr\{\hat\vm\neq (i,a,j,b)|\vx_{a}^i,\vx_{b}^j  \textnormal{\, are sent}\}.
\end{equation}
The average collision declaration error probability $Err_c (i,j)$ is defined by
\begin{equation}
\frac{1}{N_1^{i}N_2^j}\sum_{a=1}^{|A^i|}\sum_{b=1}^{|B^j|} Pr\{\hat\vm\neq \textnormal{"collision"}|\vx_{a}^i,\vx_{b}^j  \textnormal{\, are sent}\}.
\end{equation}
\end{Def}
To state our main theorem we need the following notions; the index $LH$ refers to Liu and Hughes, the authors of \cite{Hughes}.
\begin{align}
&\mathcal{V}_{LH}=\mathcal{V}_{LH}(P_{U}, P_{X|U},P_{Y|U}) \triangleq \left\{\hspace{-4pt}\begin{array}{l}
V_{UXYZ} \in \mathcal{P}(\mathcal{U} \times \mathcal{X} \times \mathcal{Y} \times \mathcal{Z}):\\ V_{UX}=P_UP_{X|U},V_{UY}=P_UP_{Y|U} \} \label{hl1}\\
\end{array}\hspace{-4pt}\right\}, \\
&\mathcal{E}X_{LH}(R_1,W,P_{U},P_{X|U},P_{Y|U}) \triangleq \notag\\
&\triangleq \min_{V_{UXYZ}  \in \mathcal{V}_{LH}}[ \DD(V_{Z|XYU} ||W|V_{XYU}) +\I_{V}(X \wedge Y|U) + |\I_{V}(X \wedge YZ|U)-R_1|^{+}],  \label{hl2} \\
&\mathcal{E}Y_{LH}(R_2,W,P_{U},P_{X|U},P_{Y|U}) \triangleq \notag\\
&\triangleq \min_{V_{UXYZ}  \in \mathcal{V}_{LH}} [\DD(V_{Z|XYU} ||W|V_{XYU}) +\I_{V}(X \wedge Y|U) + |\I_{V}(Y \wedge XZ|U)-R_2|^{+}],  \label{hl3}
\end{align}
\begin{align}
&\mathcal{E}XY_{LH}(R_1,R_2,W,P_{U}, P_{X|U},P_{Y|U}) \triangleq \notag\\
&\triangleq\min_{V_{UXYZ}  \in \mathcal{V}_{LH}} [\DD(V_{Z|XYU} ||W|V_{XYU}) +\I_{V}(X \wedge Y|U) +|\I_{V}(X \wedge Y \wedge Z|U) -R_1-R_2|^{+}],  \label{hl4} \\
&\mathcal{E}_{LH}(R_1,R_2,W,P_{U},P_{X|U},P_{Y|U}) \triangleq\min \{ \mathcal{E}X_{LH}, \mathcal{E}Y_{LH},\mathcal{E}XY_{LH} \}. \label{hl5}
\end{align}
Theorem \ref{collision} shows that the error exponent of \cite{Hughes} for an individual codebook pair is achievable for this general setting, also guaranteeing that the probability of collision declaration error goes to $0$ when it is required.

\begin{Thm} \label{collision}
For each $n$ let constant composition random access codebook library parameters as in Definition \ref{constantcomposition} be given with a common set $\mathcal{U}$, and with $\frac{1}{n}log M_1 \rightarrow 0$, $\frac{1}{n}log M_2 \rightarrow 0$ as $n \rightarrow \infty$. Then there exist a sequence $\delta_n(|\mathcal{U}|,|\mathcal{X}|,|\mathcal{Y}|,|\mathcal{Z}|,M_1,M_2) \rightarrow 0$ and for each $n$ a constant composition codebook-library pair $(\mathcal{A} ,\mathcal{B})$ with the given parameters, and decoder mappings with the following properties:
\begin{itemize}
\item[(i)] For all $(i,j) \in [M_1] \times [M_2]$
\begin{equation}
Err_d(i,j) \le 2^{-n ( \mathcal{E}_{LH}(R_1^i,R_2^j,W,P_{U},P^{i}_{X|U},P^{j}_{Y|U})-\delta_n ) }. \label{collisionbelso}
\end{equation}
\item[(ii)] If $(R_1^{i},R_2^{j})$ is not in the interior of $C[W,P_{U},P^{i}_{X|U},P^{j}_{Y|U}]$ then
\begin{equation} \label{collisionbelso2}
Err_c(i,j) < \delta_n.
\end{equation}
\end{itemize}
\end{Thm}
\begin{Rem}
The exponent $\mathcal{E}_{LH}(R_1^i,R_2^j,W,P_{U},P^{i}_{X|U},P^{j}_{Y|U})$ in part (i) of Theorem \ref{collision} is positive iff $(R_1^{i},R_2^{j})$ is in the interior of $C[W,P_{U},P^{i}_{X|U},P^{j}_{Y|U}]$.
\end{Rem}

The next packing lemma is an extension of Lemma 4 in \cite{Nazari} for this multiple codebooks setting, it provides the appropriate codebook library pair for Theorem \ref{collision}.
\begin{Lem}\label{Rc-Packing-lemma}
Let a sequence of constant composition random access codebook-library parameters be given as in Theorem \ref{collision}. Then there exist a sequence $\delta^{'}_n (|\mathcal{U}|,|\mathcal{X}|,|\mathcal{Y}|,M_1,M_2) \rightarrow 0$ and for each $n$ a constant composition codebook-library pair $(\mathcal{A},\mathcal{B})$ with the given parameters such that for any $(i,k)\in [M_1]^2$ and $(j,l)\in[M_2 ]^2$ and for all $V_{U X\hat X Y\hat Y}\in \iP^n(\iU \times \iX\times \iX \times \iY \times \iY)$:
\begin{align}
  K^{i,j}[V_{UXY}] &\leq 2^{-n(\I_{V}(X \wedge Y|U)-R_1^i-R_2^j-\delta^{'}_n)},\label{fopack1}\\
  K^{i,j}_{k}[V_{UX\hat XY}] &\leq 2^{-n(\I_{V}(X \wedge \hat X \wedge Y|U)-R_1^i-R_2^j-R_1^k-\delta^{'}_n)},\label{fopack2}\\
  K^{i,j}_{l}[V_{UXY\hat Y}] &\leq 2^{-n(\I_{V}(X \wedge Y \wedge \hat Y|U)-R_1^i-R_2^j-R_2^l-\delta^{'}_n)},\label{fopack3}\\
  K^{i,j}_{k,l}[V_{UX\hat XY\hat Y}] &\leq 2^{-n(\I_{V}(X \wedge \hat X\wedge Y \wedge \hat Y|U)-R_1^i-R_2^j-R_1^k-R_2^l-\delta^{'}_n)}, \label{fopack4}
\end{align}
where
\begin{align}
&K^{i,j}[V_{UXY}] \triangleq  \sum_{a=1}^{N_1^i}\sum_{b=1}^{N_2^j} \mathds 1_{T^{n}_{V_{UXY}}}(\vu,\vx_{a}^{i},\vy_{b}^{j}), \label{packing1} \\
&K^{i,j}_{k}[V_{UX\hat XY}] \triangleq  \sum_{a=1}^{N_1^i}\sum_{b=1}^{N_2^j}\hspace{-5pt}\sum^{N_1^k}_{\sumfrac{c=1}{c \ne a \textnormal{ if } i=k}}\hspace{-11pt}\mathds 1_{T^{n}_{V_{UX\hat XY}}}(\vu, \vx_a^i, \vx_{c}^k,\vy_{b}^j), \label{packing2} \end{align}
\begin{align}
&K^{i,j}_{l}[V_{UXY\hat Y}] \triangleq  \sum_{a=1}^{N_1^i}\sum_{b=1}^{N_2^j}\hspace{-5pt}\sum^{N_2^l}_{\sumfrac{d=1}{d \ne b \textnormal{ if } j=l}}\hspace{-11pt}\mathds 1_{T^{n}_{V_{UXY\hat Y}}}(\vu, \vx_a^i,\vy_{b}^j, \vy_{d}^l), \label{packing3}
\end{align}
\begin{align}
&K^{i,j}_{k,l}[V_{UX\hat XY\hat Y}] \triangleq \sum_{a=1}^{N_1^i}\sum_{b=1}^{N_2^j}\hspace{-4pt}\sum^{N_1^k}_{\sumfrac{c=1}{c \ne a \textnormal{ if } i = k}}\sum^{N_2^l}_{\sumfrac{d=1}{d \ne b \textnormal{ if } j = l}}\mathds 1_{T^{n}_{V_{UX \hat X Y\hat Y}}}(\vu, \vx_a^i, \vx_{c}^k,\vy_{b}^j, \vy_{d}^l). \label{packing4}
\end{align}
Here the subscripts  '$c \ne a \textnormal{ if } i=k$' and '$d \ne b \textnormal{ if } j = l$' mean that in case of $i=k$ respectively $j=l$ we do not include in the sums the terms corresponding to indices $a$ and $b$ respectively.
\end{Lem}
\begin{Rem}
Actually, (\ref{packing1})-(\ref{packing4}) are equal to $0$ if $V$ does not fulfill marginal conditions determined by the prescribed conditional distributions for the codebooks.
\end{Rem}
\begin{IEEEproof}[Proof]
Choose a constant composition codebook library pair $(\mathcal{A},\mathcal{B})$ at random, i. e, for all $i \in [M_1], j \in [M_2]$ the codewords of $A^i,B^j$ are chosen independently and uniformly from $T^{n}_{P_{X|U}^{i}}(\vu)$ and $T^{n}_{P_{Y|U}^{j}}(\vu)$ respectively. We first claim that under this random selection the expected values of the expressions of form
\begin{align}
&K^{i,j}[V_{UXY}] 2^{n(\I_{V}(X \wedge Y|U)-R_1^i-R_2^j)}, \label{pack1} \\
&K^{i,j}_{k}[V_{UX\hat XY}]2^{n(\I_{V}(X \wedge \hat X \wedge Y|U)-R_1^i-R_2^j-R_1^k)}, \label{pack2} \\
&K^{i,j}_{l}[V_{UXY\hat Y}]2^{n(\I_{V}(X \wedge Y \wedge \hat Y|U)-R_1^i-R_2^j-R_2^l)}, \label{pack3} \\
&K^{i,j}_{k,l} [V_{UX\hat XY\hat Y}] 2^{n(\I_{V}(X \wedge \hat X\wedge Y \wedge \hat Y|U)-R_1^i-R_2^j-R_1^k)} \label{pack4}
\end{align}
are bounded above by a polynomial of $n$ that depends only on the alphabet sizes $|\mathcal{U}|$, $|\mathcal{X}|$, $|\mathcal{Y}|$. We establish this claim for (\ref{pack4}), the other cases are similar. Now, if $(i,a) \ne (k,c)$ and $(j,b) \ne (l,d)$ then
\begin{equation}
   \EE(\mathds 1_{T^{n}_{V_{UX \hat X Y\hat Y}}}(\vu, \vX_a^i, \vX_{c}^k,\vY_{b}^j, \vY_{d}^l)= \hspace{-5pt}\sum_{\sumfrac{(\vx,\hat\vx,\vy,\hat\vy)\in}{T^n_{V_{X\hat XY\hat Y|U}}(\vu)}}\hspace{-20pt}P(\vX_a^i=\vx)P(\vX_{c}^k=\hat\vx)P(\vY_{b}^j=\vy)P(\vY_{d}^l=\hat\vy), \label{pack5}
\end{equation}	
where $\vX_a^i, \vX_{c}^k,\vY_{b}^j, \vY_{d}^l$ denote the random codewords. Using (\ref{basicfact3}) and (\ref{ujjeloles}), the sum in (\ref{pack5}) is bounded above by
\begin{equation} \label{pack6}
2^{-n(\I_{V}(X \wedge \hat X\wedge Y \wedge \hat Y|U))} (n+1)^{2|\mathcal{U}|(|\mathcal{X}|+|\mathcal{Y}|)}.
\end{equation}
Recalling the definition of $N_1^i$ and $N_2^j$ in Definition \ref{constantcomposition}, this establishes the claim for (\ref{pack4}).

Next, denote by $S$ the sum of the expressions of form (\ref{pack1})-(\ref{pack4}) for all $i$, $j$, $k$, $l$ and joint type $V_{UX\hat{X}Y\hat{Y}}$. As $M_1$, $M_2$ grow at most subexponentially with $n$, and the number of joint types is polynomial, it follows that $\EE[S] \le 2^{n \delta^{'}_n}$ for suitable $\delta^{'}_n (|\mathcal{U}|,|\mathcal{X}|,|\mathcal{Y}|,M_1,M_2) \rightarrow 0$. Hence, there exists a realization of the codebook library pair with $S \le 2^{n \delta^{'}_n}$. Then, clearly, inequalities (\ref{fopack1})-(\ref{fopack4}) hold. \end{IEEEproof}

\begin{IEEEproof}[Proof of Theorem \ref{collision}]
Lemma \ref{Rc-Packing-lemma} provides the appropriate constant composition codebook-library pair $(\mathcal{A},\mathcal{B})$. To construct the decoder, define $\alpha: \iP (\mathcal{U} \times \mathcal{X} \times \mathcal{Y} \times \mathcal{Z}) \rightarrow \mathbb{R}$ by $\alpha (V_{UXYZ})=\I_{V}(X \wedge Y \wedge Z | U )$. In the first stage of decoding, the receiver tries to find indices $\hat{k} \in [M_1]$, $\hat{c} \in [N_1^{\hat{k}}]$, $\hat{l} \in [M_2]$, $\hat{d} \in [N_2^{\hat{l}}]$ which uniquely maximize
\begin{equation}
\alpha (\vu,\vx_{c}^k,\vy_{d}^{l},\vz) - R_1^{k} - R_2^{l}, \label{maximum}
\end{equation}
where $\vz$ denotes the output sequence and $\alpha$ is evaluated on the joint type of $(\vu,\vx_{c}^k,\vy_{d}^{l},\vz)$. If the decoder succesfully finds a unique maximizer $(\hat{k},\hat{c},\hat{l},\hat{d})$, the second stage of decoding starts. In this stage, to deal with collisions, the decoder checks the following three inequalities:
\begin{align}
&\I(\vx_{\hat{c}}^{\hat{k}} \wedge \vy_{\hat{d}}^{\hat{l}} \wedge \vz | \vu) -  R_1^{\hat{k}} - R_2^{\hat{l}} > \eta_n, \label{crit1} \\
&\I(\vx_{\hat{c}}^{\hat{k}} \wedge \vy_{\hat{d}}^{\hat{l}},\vz | \vu) -  R_1^{\hat{k}} > \eta_n, \label{crit2} \\
&\I(\vy_{\hat{d}}^{\hat{l}} \wedge \vx_{\hat{c}}^{\hat{k}},\vz | \vu) - R_2^{\hat{l}} > \eta_n, \label{crit3}
\end{align}
where $\eta_n(|\mathcal{U}|,|\mathcal{X}|,|\mathcal{Y}|,|\mathcal{Z}|,M_1, M_2) \rightarrow 0$ is an appropriately chosen positive sequence. If the above three inequalities are fulfilled then the decoder decodes $\vx_{\hat{c}}^{\hat{k}}$, $\vy_{\hat{d}}^{\hat{l}}$ as the codewords sent, if at least one of them is not fulfilled, then the decoder reports ``collision''.

The necessary calculations can be found in Appendix \ref{bizonyitas}.
\end{IEEEproof}
\begin{Rem}
Using threshold criterion (\ref{crit1})-(\ref{crit3}) is a key idea of this paper. A somewhat similar idea for one-way channels appears in the proof of Theorem 1 in \cite{dispersioncikk}. Note also that this threshold criterion is essential for collision detection, but is not necessary for achieving part (i) of Theorem \ref{collision}.
\end{Rem}
\begin{Rem}  \label{nonuniversal2}
Appendix \ref{elesseg} shows that if the universal decoder in the proof of Theorem \ref{collision} is used, the collision error probability may not go to zero exponentially fast. Nevertheless, it is possible to modify this decoder by increasing the threshold $\eta_n$ to some positive constant. This could give rise to a collision error exponent at the expense of decreasing the exponent of decoding error and possibly leading to collision declaration also for rate pairs inside their pentagons. Using a constant positive threshold appears reasonable mainly if the receiver knows the channel.
\end{Rem}
\begin{Rem}
When both senders have only one codebook, then maximizing $\I  (\vx_{c} \wedge \vy_{d} \wedge \vz|\vu) - R_1 - R_2$ is equivalent to minimizing $\HH(\vx_{c},\vy_{d} | \vu, \vz)$ which was the decoder of \cite{Hughes}.
\end{Rem}
\begin{Rem} \label{nonuniversal}
Other $\alpha$-decoders can also be used (but could be more difficult to analyze); if the receiver knows the channel $W$, the function $\alpha$ can depend on $W$. For the sake of brevity, the expurgation method for multiple access channel in \cite{Nazari} is not used in this paper. However, it is possible to prove an expurgated version of Lemma \ref{Rc-Packing-lemma} which yields larger achievable error exponent for small rates.
\end{Rem}

\section{Error exponents in source-channel coding} \label{sc}
Let two independent discrete memoryless sources (DMS) $Q_1$, $Q_2$ with alphabets $\iS_1$, $\iS_2$ be given. We want to transmit these sources over MAC $W$. We apply the results of Section \ref{Modell} to get achievable error exponent in classical source-channel coding setting and in a slightly modified setting where a special error free $0$ rate communication is allowed between the senders. A standing assumption in this section is that $(H(Q_1),H(Q_2))$ is in the interior of $C(W)$, hence, the sources can be reliably transmitted over channel $W$.

\subsection{Classical setting} \label{scc}
\begin{Def}
A source-channel code of length $n$ is a mapping triple $(f_1,f_2, \varphi)$ with encoders $f_1: \iS_1^n \rightarrow \mathcal{X}^n$, $f_2: \iS_2^n \rightarrow \mathcal{Y}^n$ and decoder $\varphi: \mathcal{Z}^n \rightarrow \iS_1^n \times \iS_2^n$.
\end{Def}
\begin{Def}
The error probability of a source-channel code $(f_1,f_2, \varphi)$ of length $n$ is defined by
\begin{align}
&Err(f_1,f_2, \varphi)= \sum_{\sumfrac{(\mathbf{s}_1,\mathbf{s}_2)\in}{ \iS_1^n \times \iS_2^n}} Q^n_{1}(\mathbf{s}_1)Q^n_{2}(\mathbf{s}_2) p_e (\mathbf{s}_1,\mathbf{s}_2), \text{ where} \\
&p_e (\mathbf{s}_1,\mathbf{s}_2) =W^n (\{\vz \in \mathcal{Z}^n: \varphi(\vz)\ne (\mathbf{s}_1,\mathbf{s}_2)  \} | f_1 (\mathbf{s}_1),f_2 (\mathbf{s}_2)).
\end{align}
\end{Def}
\begin{Def} \label{exponensdef}
A number $\mathcal{E} > 0$ is called achievable error exponent for transmission of sources $Q_1$, $Q_2$ over channel $W$ if to every $\delta >0$ and for every sufficiently large $n$ there exists source-channel code $(f_1,f_2, \varphi)$ of length $n$ with $Err(f_1,f_2, \varphi)$ less than $2^{-n(\mathcal{E}-\delta)}$.
\end{Def}
For arbitrary $\mathcal{U}$ let $G_{1}(\mathcal{U})$ and $G_{2}(\mathcal{U})$ be the set of all mappings $[0,\log|S_1|] \rightarrow \iP (\iX|\mathcal{U})$ and $[0,\log|S_2|] \rightarrow \iP (\iY|\iU)$ respectively, and define
\begin{equation} \label{ejexponens}
\mathcal{E}j(Q_1,Q_2,W) \triangleq\hspace{-6pt} \sup_{\sumfrac{\mathcal{U}}{P_{U} \in \mathcal{P} (\mathcal{U})}} \sup_{\sumfrac{g_{1} \in G_{1} (\iU)}{g_{2} \in G_{2} (\iU)}}\hspace{-4pt} \mathcal{E}j(Q_1,Q_2,W,P_U,g_1,g_2)
\end{equation}
where
\begin{equation}
\mathcal{E}j(Q_1,Q_2,W,P_U,g_1,g_2) \triangleq \min_{\sumfrac{0 \le R_1 \le \log|S_1|}{0 \le R_2 \le \log|S_2|}}\Big[e_{1}(R_1, Q_1) + e_{2}(R_2, Q_2) +\mathcal{E}_{LH}(R_1,R_2,W,P_{U},g_1 (R_1),g_2 (R_2))\Big] \label{ej},
\end{equation}
and $e_{1}(R_1,Q_1)$, $e_{2}(R_2,Q_2)$ are the source reliability functions
\begin{equation}
e_{i}(R_i,Q_i) \triangleq \min_{P:\HH(P)\geq R_i}\DD(P\|Q_i),\quad i \in \{1,2\}. \label{rel}
\end{equation}

The following theorem shows that $\mathcal{E}j(Q_1,Q_2,W)$ is an achievable error exponent for this source-channel coding scenario. More exactly, we show that for any choice of $P_U$, $g_1$, $g_2$, the exponent $\mathcal{E}j(Q_1,Q_2,W,P_U,g_1,g_2)$ is achievable even if the senders and the receiver do not know the sources and the channel; if they do know them, they can optimize in $P_U$, $g_1$, $g_2$, to achieve $\mathcal{E}j(Q_1,Q_2,W)$ (the suprerum of achievable error exponents is an achievable error exponent). Note that in both cases the alphabets $\mathcal{S}_1$, $\mathcal{S}_2$, $\mathcal{X}$, $\mathcal{Y}$, $\mathcal{Z}$ are assumed to be known by the senders and the receiver.
\begin{Thm} \label{source-channel}
Let $\mathcal{U}$, $P_{U} \in \mathcal{P} (\mathcal{U})$, $g_1 \in G_{1} (\mathcal{U})$ and $g_2 \in G_{2} (\mathcal{U})$ be given. There exist a source-channel code for each $n$ and a sequence $\nu_n (|\mathcal{S}_{1}|,|\mathcal{S}_{2}|,|\mathcal{U}|,|\mathcal{X}|,|\mathcal{Y}|,|\mathcal{Z}|) \rightarrow 0$  with
\begin{equation}
Err(f_1,f_2, \varphi) \le 2^{ -n(\mathcal{E}j(Q_1,Q_2,W,P_U,g_1,g_2)-\nu_n)}.
\end{equation}
\end{Thm}
\begin{IEEEproof}
Approximate\footnote{We use here the fact that distributions on a finite set can be uniformly approximated by types.} $P_U$, $g_1$, $g_2$ by $P_U[n] \in \mathcal{P}^n (\mathcal{U})$, $g_1 [n]: [0,\log|S_1|] \rightarrow \iP^n (\iX|P_U (n))$, $g_2 [n]: [0,\log|S_2|] \rightarrow \iP^n (\iY|P_U (n))$ such that
\begin{equation}
\sup_{R_i \in [0,\log|S_i|]} ||P_U g_i(R_i)-P_U[n]g_i [n] (R_i)|| \le \nu^{'}_n, \label{approx2}
\end{equation}
with $\nu^{'}_n(|\mathcal{U}|,|\mathcal{X}|,|\mathcal{Y}|) \rightarrow 0$, $\text{ } i \text{ } \in \text{ } \{1,2\}$.

Let $\mathbf{u} \in T^n_{P_{U}[n]}$ be arbitrary. Let $P_1^1,P_1^2, \dots, P_1 ^ {|\mathcal{P}^n (S_1)|}$ and $P_2^1,P_2^2, \dots, P_2 ^ {|\mathcal{P}^n (S_2)|}$ denote all possible types from $\mathcal{P}^n(\mathcal{S}_1)$ and $\mathcal{P}^n(\mathcal{S}_2)$ respectively. Choose $M_1 = |\mathcal{P}^n (S_1)|$ and $M_2  = |\mathcal{P}^n (S_2)|$. For all $i \in [M_1 ]$, $j \in [M_2]$ let $R_1 ^i$ and $R_2 ^j$ be equal to $\frac{1}{n}\log|T^{n}_{P_1^i}|$ and $\frac{1}{n}\log|T^{n}_{P_2^j}|$ respectively, and let $P^i_{X|U}$ and $P^j_{Y|U}$ be equal to $g_1[n](R_1 ^i)$, $g_2[n](R_2 ^j)$ respectively. Applying Theorem \ref{collision} with these parameters consider the resulting codebook library pair $(\mathcal{A},\mathcal{B})$ and the decoder mapping $\phi$ satisfying (\ref{collisionbelso}) for all $(i,j) \in [M_1] \times [M_2]$.

Let $f_1: \mathcal{S}_1^n \rightarrow \mathcal{X}^n$ and $f_2: \mathcal{S}_2^n \rightarrow \mathcal{Y}^n$ be mappings which map each $T^{n}_{P_1^i}$ and $T^{n}_{P_2^j}$ to $A^i$ and $B^j$ respectively. Let $\varphi: \mathcal{Z}^n \rightarrow \mathcal{S}_1^n \times \mathcal{S}_2^n$ be the mapping which first determines a codeword pair from $(\mathcal{A},\mathcal{B})$ using $\phi$, then uses the inverse of $f_1$ and $f_2$ to determine the source sequences. The crucial step is the following equation
\begin{align}
Err(f_1,f_2, \varphi)= &\sum_{i=1}^{|\mathcal{P}^n (S_1)|}\sum_{j=1}^{|\mathcal{P}^n (S_2)|} Q_1^n (T^{n}_{{P_1^i}})Q_2^n (T^{n}_{{P_2^j}}) \cdot \notag\\
 &\cdot\frac{1}{|T^{n}_{{P_1^i}}|}\frac{1}{|T^{n}_{{P_2^j}}|} \sum_{\mathbf{s}_1 \in T^{n}_{{P_1^i}}}\sum_{\mathbf{s}_2 \in T^{n}_{{P_2^j}}} p_e (\mathbf{s}_1,\mathbf{s}_2). \label{felsobecsles3}
\end{align}
Note that the second line of (\ref{felsobecsles3}) is $Err_d (i,j)$ in the termin\-o\-logy of Theorem \ref{collision}. Hence substituting (\ref{collisionbelso}) into (\ref{felsobecsles3}) and using (\ref{basicfact2}), (\ref{rel}), and (\ref{approx2}) with the continuity of the present information measures, this theorem is proved. \end{IEEEproof}

To analyze the achieved exponent it is useful to consider the following expression:
\begin{equation}
\mathcal{E}s_{LH}(Q_1,Q_2,W) \triangleq \max_{\sumfrac{0 \le R_1 \le \log|S_1|}{0 \le R_2 \le \log|S_2|}} \min\Big[e_{1}(R_1,Q_1),e_{2}(R_2,Q_2),\mathcal{E}_{LH}(R_1,R_2,W)\Big], \label{eshl}
\end{equation}
where
\begin{equation}
\mathcal{E}_{LH}(R_1,R_2,W) \triangleq\hspace{-6pt}\sup_{\sumfrac{\mathcal{U}}{P_U \in \mathcal{P}(\mathcal{U})}} \sup_{\sumfrac{P_{X|U} \in \iP (\iX|\iU)}{P_{Y|U \in \iP (\iY|\iU)}}}\hspace{-10pt} \mathcal{E}_{LH}[R_1,R_2,W,P_U,P_{X|U},P_{Y|U}].\label{esh2}
\end{equation}
\begin{Rem} \label{korlatoz}
In \cite{Hughes} it is proved that the suprerum in (\ref{esh2}) is attained with $|\mathcal{U}|=4$.
\end{Rem}
Note that $\mathcal{E}s_{LH} (Q_1,Q_2,W)$ was introduced in \cite{mi} as an exponent achievable by separate source and channel coding, which can be improved replacing $\mathcal{E}_{LH}(R_1,R_2,W)$ by the reliability function of channel $W$. This statement, however, has turned out not so easy to prove. The main difficulty is that the channel coding exponent holds for average (rather than maximal) error, and the classical source coding schemes yield non-uniformly distributed output (see, for example, \cite{remi}). Even the proper definition of "separate source-channel coding" is not particularly obvious. In \cite{dispersioncikk} a possible definition is given (for one-way channels, but easily extendable to multiple access channels) which circumvents the mentioned difficulty using shared randomness. Under that definition, $\mathcal{E}s_{LH} (Q_1,Q_2,W)$ is certainly achievable by separate source-channel coding, but we believe this should also be true under a more appealing definition. This question, through of interest in itself, is a side issue for this paper. The emphasis is on the fact that $\mathcal{E}s_{LH} (Q_1,Q_2,W)>0$ and on the following proposition.
\begin{all} \label{dominal}
$\mathcal{E}j(Q_1,Q_2,W) \ge \mathcal{E}s_{LH}(Q_1,Q_2,W)$.
\end{all}
\begin{IEEEproof}
Restricting the suprerum to constant functions $g_1$, $g_2$ in (\ref{ejexponens}), $\mathcal{E}j(Q_1,Q_2,W)$ can be lower-bonded by
\begin{equation}
\sup_{\sumfrac{\mathcal{U}}{P_U \in \mathcal{P}(\mathcal{U})}} \sup_{\sumfrac{P_{X|U} \in \iP (\iX|\iU)}{P_{Y|U \in \iP (\iY|\iU)}}} \min_{\sumfrac{0 \le R_1 \le log|S_1|}{0 \le R_2 \le log|S_2|}}\Big[e_{1}(R_1, Q_1) +e_{2}(R_2,Q_2)+\mathcal{E}_{LH}(R_1,R_2,W,P_U,P_{X|U},P_{Y|U})\Big]. \label{alsobecsles1}
\end{equation}
As a consequence of Remark \ref{korlatoz} the suprerum is attained in (\ref{esh2}), hence, there exist $R_1^{*}$, $R_2^{*}$, $P_U^{*}$, $P_{X|U}^{*}$ and $P_{Y|U}^{*}$ with
\begin{equation} \label{alsobecsles2}
\mathcal{E}s_{LH}(Q_1,Q_2,W)=\min\Big[e_{1}(R^{*}_1,Q_1), e_{2}(R^{*}_2,Q_2),\mathcal{E}_{LH}(R^{*}_1,R^{*}_2,W,P^{*}_U,P^{*}_{X|U},P^{*}_{Y|U})\Big].
\end{equation}
Using this notation (\ref{alsobecsles1}) can be lower-bounded by
\begin{equation}
\min_{\sumfrac{0 \le R_1 \le log|S_1|}{0 \le R_2 \le log|S_2|}}\Big[e_{1}(R_1, Q_1)+e_{2}(R_2,Q_2)+\mathcal{E}_{LH}(R_1,R_2,W,P^{*}_U,P^{*}_{X|U},P^{*}_{Y|U})\Big]. \label{alsobecsles3}
\end{equation}
Using the definition of $e_{1}(R_1, Q_1)$, $e_{2}(R_2,Q_2)$ and $\mathcal{E}_{LH}(R_1,R_2,W,P_{U},P_{X|U},P_{Y|U})]$ it can be seen that for any $(R_1,R_2)$ with $0 \le R_1 \le log|S_1|$ and $0 \le R_2 \le log|S_2|$ it is true that $e_{1}(R_1, Q_1) \ge e_{1}(R^{*}_1, Q_1)$ or $e_{2}(R_2, Q_2) \ge e_{2}(R^{*}_2, Q_2)$ or $\mathcal{E}_{LH}(R_1,R_2,W,P^{*}_U,P^{*}_{X|U},P^{*}_{Y|U}) \ge \mathcal{E}_{LH}(R^{*}_1,R^{*}_2,W,P^{*}_U,P^{*}_{X|U},P^{*}_{Y|U})$. Via this fact and (\ref{alsobecsles1})-(\ref{alsobecsles3}) the proposition is proved.
\end{IEEEproof}
\begin{Rem}
In Proposition \ref{dominal} the inequality is strict except in very special cases.
\end{Rem}

\subsection{Special $0$ rate communication is allowed between senders}  \label{sc0}
In this subsection we allow error free $0$ rate communication between the senders. More exactly, as the following definition shows, it is assumed that the senders are able to inform each other about the types of their source sequences.
\begin{Def}
A type-informed source-channel code of length $n$ is a mapping triple $(f_1,f_2, \varphi)$ with encoders $f_1: \iS_1^n \times \mathcal{P}^n (S_2) \rightarrow \mathcal{X}^n$, $f_2: \iS_2^n \times \mathcal{P}^n (S_1) \rightarrow \mathcal{Y}^n$ and decoder $\varphi: \mathcal{Z}^n \rightarrow \iS_1^n \times \iS_2^n$.
\end{Def}
\begin{Def}
The error probability of a type-informed source-channel code $(f_1,f_2, \varphi)$ of length $n$ is defined by
\begin{align}
&Err_{0}(f_1,f_2, \varphi)= \sum_{\sumfrac{(\mathbf{s}_1,\mathbf{s}_2)\in}{ \iS_1^n \times \iS_2^n}} Q^n_{1}(\mathbf{s}_1)Q^n_{2}(\mathbf{s}_2) p_{e_{0}} (\mathbf{s}_1,\mathbf{s}_2), \\
&\text{ where } p_{e_{0}} (\mathbf{s}_1,\mathbf{s}_2)=W^n (\{\vz \in \mathcal{Z}^n: \varphi(\vz)\ne (\mathbf{s}_1,\mathbf{s}_2)  \} | f_1 (\mathbf{s}_1, P_{\mathbf{s}_2}),f_2 (\mathbf{s}_2, P_{\mathbf{s}_1})).
\end{align}
\end{Def}
\begin{Def}
A number $\mathcal{E}_{0} > 0$ is called achievable error exponent for type-informed transmission of sources $Q_1$, $Q_2$ over channel $W$ if to every $\delta >0$ and for every sufficiently large $n$ there exists type-informed source-channel code $(f_1,f_2, \varphi)$ of length $n$ with $Err_{0}(f_1,f_2, \varphi)$ less than $2^{-n(\mathcal{E}_{0}-\delta)}$.
\end{Def}
For arbitrary $\mathcal{U}$ let $G_{0,1}(\mathcal{U})$ and $G_{0,2}(\mathcal{U})$ be the set of all mappings $[0,\log|S_1|] \times [0,\log|S_2|] \rightarrow \iP (\iX|\mathcal{U})$ and $[0,\log|S_1|] \times [0,\log|S_2|] \rightarrow \iP (\iY|\iU)$ respectively, and define $\mathcal{E}j_{0}(Q_1,Q_2,W)$ by
\begin{equation} \label{ej0exponens}
\sup_{\sumfrac{\mathcal{U}}{P_{U} \in \mathcal{P} (\mathcal{U})}} \sup_{\sumfrac{g_1 \in G_{0,1}(\mathcal{U})}{g_2 \in G_{0,2}(\mathcal{U})}}\hspace{-4pt} \mathcal{E}j_{0}(Q_1,Q_2,W,P_U,g_1,g_2)
\end{equation}
where
\begin{equation}
\mathcal{E}j_{0}(Q_1,Q_2,W,P_U,g_1,g_2) \triangleq \min_{\sumfrac{0 \le R_1 \le \log|S_1|}{0 \le R_2 \le \log|S_2|}}\Big[e_{1}(R_1, Q_1) + e_{2}(R_2, Q_2) +\mathcal{E}_{LH}(R_1,R_2,W,P_{U},g_1(R_1,R_2),g_2(R_1,R_2))\Big] \label{ej0}.
\end{equation}
\begin{Rem}
In (\ref{ej0exponens}) we take suprerum over a larger set than in (\ref{ejexponens}), hence,  $\mathcal{E}j_{0}(Q_1,Q_2,W) \ge \mathcal{E}j(Q_1,Q_2,W)$.
\end{Rem}

The following theorem shows that for any choice of $P_U$, $g_1$, $g_2$ the exponent $\mathcal{E}j_{0}(Q_1,Q_2,W,P_U,g_1,g_2)$ is achievable even if the senders and the receiver do not know the sources and the channel (the alphabets are known); if they do know them, they can optimize in $P_U$, $g_1$ and $g_2$ to achieve $\mathcal{E}j_{0}(Q_1,Q_2,W)$.
\begin{Thm} \label{source-channel0}
Let $\mathcal{U}$, $P_{U} \in \mathcal{P} (\mathcal{U})$, $g_1 \in G_{0,1} (\mathcal{U})$ and $g_2 \in G_{0,2} (\mathcal{U})$ be given. There exist a type-informed source-channel code for each $n$ and a sequence $\mu_n (|\mathcal{S}_{1}|,|\mathcal{S}_{2}|,|\mathcal{U}|,|\mathcal{X}|,|\mathcal{Y}|,|\mathcal{Z}|) \rightarrow 0$ with
\begin{equation}
Err_{0}(f_1,f_2, \varphi) \le 2^{ -n(\mathcal{E}j_{0}(Q_1,Q_2,W,P_U,g_1,g_2)-\mu_n)}.
\end{equation}
\end{Thm}\begin{IEEEproof}
Approximate unformly in variational distance (similarly as in (\ref{approx2})) $P_U$, $g_1$, $g_2$ by $P_U[n] \in \mathcal{P}^n (\mathcal{U})$, $g_1 [n]: [0,\log|S_1|] \times [0,\log|S_2|] \rightarrow \iP^n (\iX|P_U (n))$, $g_2 [n]: [0,\log|S_1|] \times [0,\log|S_2|] \rightarrow \iP^n (\iY|P_U (n))$.

Let $P_1^1,P_1^2, \dots, P_1 ^ {|\mathcal{P}^n (S_1)|}$ and $P_2^1,P_2^2, \dots, P_2 ^ {|\mathcal{P}^n (S_2)|}$ denote all possible types from $\mathcal{P}^n(\mathcal{S}_1)$ and $\mathcal{P}^n(\mathcal{S}_2)$ respectively. We use Theorem \ref{collision} with the following parameters. Let $\mathbf{u} \in T^n_{P_{U}[n]}$ be arbitrary. The senders have the same number of codebooks in their codebook libraries: $M_1=M_2=|\mathcal{P}^n (S_1)| |\mathcal{P}^n (S_2)|$. We index the codebooks in the codebook libraries by pairs $(k,l)$, $k \in [|\mathcal{P}^n (S_1)|]$, $l \in [|\mathcal{P}^n (S_2)|]$. For all possible indices $(k,l)$ let $R_1 ^{(k,l)}$ and $R_2 ^{(k,l)}$ be equal to $\frac{1}{n}\log|T^{n}_{P_1^k}|$ and $\frac{1}{n}\log|T^{n}_{P_2^l}|$ respectively, and let $P^{(k,l)}_{X|U}$ and $P^{(k,l)}_{Y|U}$ be equal to $g_1[n](R_1 ^{(k,l)}, R_2^{(k,l)})$, $g_2[n](R_1 ^{(k,l)}, R_2^{(k,l)})$ respectively. Applying Theorem \ref{collision} with these parameters consider the resulting codebook library pair $(\mathcal{A},\mathcal{B})$ and the decoder mapping $\phi$.

Let $f_1: \iS_1^n \times \mathcal{P}^n (S_2) \rightarrow \mathcal{X}^n$ and $f_2: \iS_2^n \times \mathcal{P}^n (S_1) \rightarrow \mathcal{Y}^n$ be the mappings which map each $T^{n}_{P_1^k} \times \{P_2^l\} \subset \iS_1^n \times \mathcal{P}^n (S_2)$ and $T^{n}_{P_2^l} \times \{P_1^k\} \subset \iS_2^n \times \mathcal{P}^n (S_1)$ to $A^{(k,l)}$ and $B^{(k,l)}$ respectively. Let $\varphi: \mathcal{Z}^n \rightarrow \mathcal{S}_1^n \times \mathcal{S}_2^n$ be the mapping which first determines a codeword pair from $(\mathcal{A},\mathcal{B})$ using $\phi$, then uses the inverse of $f_1$ and $f_2$ to determine the source sequences.  The following equation finishes the proof because its second part is $Err_d ( (k,l),(k,l))$ in the termin\-o\-logy of Theorem \ref{collision} (see also the end of the proof of Theorem \ref{source-channel}).
\begin{align}
Err(f_1,f_2, \varphi)= &\sum_{k=1}^{|\mathcal{P}^n (S_1)|}\sum_{l=1}^{|\mathcal{P}^n (S_2)|} Q_1^n (T^{n}_{{P_1^k}})Q_2^n (T^{n}_{{P_2^l}})  \frac{1}{|T^{n}_{{P_1^k}}|}\frac{1}{|T^{n}_{{P_2^l}}|} \sum_{\mathbf{s}_1 \in T^{n}_{{P_1^k}}}\sum_{\mathbf{s}_2 \in T^{n}_{{P_2^l}}} p_{e_{0}} (\mathbf{s}_1,\mathbf{s}_2). \label{felsobecsles30}
\end{align} 
\end{IEEEproof}
The following theorem gives a simpler equivalent form for the achieved exponent.
\begin{Thm} \label{exponensegyenloseg}
The achievable exponent $\mathcal{E}j_{0}(Q_1,Q_2,W)$ for type-informed transmission of sources $Q_1$, $Q_2$ over channel $W$ is equal to
\begin{equation} \label{ekvivalens}
\min_{\sumfrac{0 \le R_1 \le \log|S_1|}{0 \le R_2 \le \log|S_2|}} [e_{1}(R_1, Q_1)+e_{2}(R_2,Q_2)+\mathcal{E}_{LH}(R_1,R_2,W)].
\end{equation}
\end{Thm}
\begin{IEEEproof}
Using (\ref{esh2}) the last term in (\ref{ej0}) can be upper-bounded by $\mathcal{E}_{LH}(R_1,R_2,W)$. Hence, (\ref{ekvivalens}) upper-bounds $\mathcal{E}j_{0}(Q_1,Q_2,W)$.

The other direction can be proved in the following way. Denote by $P_{U} (R_1,R_2)$, $P_{X|U} (R_1,R_2)$, $P_{Y|U} (R_1,R_2)$ the maximizing distributions in (\ref{esh2}). Using Remark \ref{korlatoz} it can be assumed that for each rate pair $(R_1,R_2)$ the cardinality of the support of $P_{U} (R_1,R_2)$ is $4$. Denote this support by $\{a,b,c,d\}$.

In (\ref{ej0exponens}) let $\mathcal{U}$ be $[k]$ and $P_U$ be the uniform distribution on this set. For each rate pair $(R_1,R_2)$ divide $[k]$ into $4$ disjoint sets $S_{\beta}$, $\beta \in \{a,b,c,d\}$ such that $P_{U} (S_{\beta})$ approximates $P_{U} (R_1,R_2) (\beta)$. For $u \in S_{\beta}$, $\beta \in \{a,b,c,d\}$, let $g_1 (R_1,R_2)$ and $g_2 (R_1,R_2)$ conditioned on $u$ be $P_{X|U=\beta} (R_1,R_2)$ and $P_{Y|U=\beta} (R_1,R_2)$ respectively. With this particular choice
\begin{equation}
\mathcal{E}_{LH}(R_1,R_2,W,P_{U},g_1(R_1,R_2),g_2(R_1,R_2))
\end{equation}
approximates $\mathcal{E}_{LH}(R_1,R_2,W)$ uniformly in $(R_1,R_2)$ as $k$ goes to infinity. This establishes the other direction.      \end{IEEEproof}
\begin{Rem}
Theorem \ref{exponensegyenloseg} provides the most direct counterpart of the JSCC result in \cite{Csiszar}. Note, howevever, that the achievability of the error exponent in Theorem \ref{exponensegyenloseg} has been proved only with type-informed encoders.
\end{Rem}
\begin{Rem}
Analogously to Lemma 2 of \cite{Csiszar}, it follows from (\ref{felsobecsles30}) that even if this special error free $0$ rate communication is allowed between the senders the error exponent cannot be greater than \begin{equation} \label{exponensfelsokorlat}
\min_{\sumfrac{0 \le R_1 \le \log|S_1|}{0 \le R_2 \le \log|S_2|}} [e_{1}(R_1, Q_1)+e_{2}(R_2,Q_2)+\mathcal{E}(R_1,R_2,W)]
\end{equation}
where $\mathcal{E}(R_1,R_2,W)]$ is the (unknown) reliability function of channel $W$. If not only this special $0$ rate communication is allowed, but the channel $W$ has positive $0$ error capacity, then (\ref{exponensfelsokorlat}) is easily achievable by communicating the types of the source sequences to the receiver with $o(n)$ symbols.
Not entering here the question whether the last assertion holds even if no communication whatsoever is allowed between senders, we only mention that we believe to have a counterexample.
\end{Rem}

\section{Conclusions} \label{conc}
A version of the random access model has been studied. We have shown that the error exponent of \cite{Hughes} for an individual codebook pair is simultaneously achievable for each codebook pair in the codebook libraries, supplemented with collision detection in the sense of \cite{Ephremides,Jie-Luo}. Moreover, this has been achieved with a universal decoder. In particular, a positive answer has been given to the question in \cite{Ephremides} whether or not the results there are still valid if the receiver does not know the channel.

The random access model has been connected with the problem of source-channel transmission of independent sources via MAC. We have given an achievable error exponent to this problem together with an improvement when error free $0$ rate communication is allowed between the senders. This improvement is the most direct counterpart of the JSCC result in \cite{Csiszar}.

In general little is known about the tightness of error exponents for MACs, and this applies for the achieved error exponents of this paper. However, for a specific class of MACs the authors have a heuristic proof that (i) the error exponent $E_{LH}(R_1,R_2,W)$ of Liu and Hughes is tight for "large" $(R_1,R_2)$ (ii) the exponent in Theorem \ref{exponensegyenloseg} is tight if the minimum in (\ref{ekvivalens}) is attained for $(R_1,R_2)$ as in (i) (iii) this exponent is not achievable if no communication is allowed between the senders. We intend to return to this issue elsewhere.

\appendices

\section{Simple identity for multi-information} \label{kislemma}
The following lemma is useful in Appendix \ref{bizonyitas}.
\begin{Lem}[Special case of Theorem 1 in \cite{watanabe}] \label{multiinfo}
Let a vector valued random variable $(X_1,X_2,\dots,X_N,U)$ be given. Let $(\mathcal{I},\mathcal{J})$ be a partition of $[N]$, $\mathcal{I}=\{i_1,i_2,\dots, i_k\}$, $\mathcal{J}= \{j_1, j_2, \dots,j_{N-k}\}$. Then $\I(X_1 \wedge X_2 \wedge \dots \wedge X_n|U)$ is equal to
\begin{equation}
\I(X_{i_1} \wedge X_{i_2} \wedge \dots \wedge X_{i_k}|U)+\I(X_{j_1} \wedge X_{j_2} \wedge \dots \wedge X_{j_{N-k}}|U) + \I(X_{i_1}, X_{i_2}, \dots,X_{i_k} \wedge X_{j_1}, X_{j_2}, \dots,X_{j_{N-k}}|U). \label{ezjonki}
\end{equation}
\end{Lem}
\begin{IEEEproof}
\begin{align}
&\I(X_1 \wedge X_2 \wedge \dots \wedge X_n|U)= \sum_{i \in \mathcal{I}} \HH(X_i|U) + \sum_{j \in \mathcal{J}} \HH(X_j|U)-\HH(X_1,X_2,\dots,X_n|U) \\
&= \Big( \sum_{i \in \mathcal{I}} \HH(X_i|U) - \HH(X_{i_1} \wedge X_{i_2} \wedge \dots \wedge X_{i_k}|U) \Big) \label{haromzarojel} \\
&+ \Big( \sum_{j \in \mathcal{J}} \HH(X_j|U) - \HH(X_{j_1} \wedge X_{j_2} \wedge \dots \wedge X_{j_{N-k}}|U) \Big) \notag\\
&+ \Big( \HH(X_{i_1} \wedge X_{i_2} \wedge \dots \wedge X_{i_k}|U)+\HH(X_{j_1} \wedge X_{j_2} \wedge \dots \wedge X_{j_{N-k}}|U)
-\HH(X_1,X_2,\dots,X_n|U) \Big). \notag
\end{align}
\end{IEEEproof}

\section{Proof of Theorem \ref{collision} (calculations)} \label{bizonyitas}
In this section we suppose that a codebook library pair as in Lemma \ref{Rc-Packing-lemma} is given, and prove rigorously that, with the decoder specified in the proof of Theorem \ref{collision}, it fulfills the error requirements.

\subsection*{Proof of part (i) of Theorem \ref{collision}}

\emph{Step 1: Some definitions}

The following definitions are meant for all possible combinations of indices $i\in [M_1]$, $j\in [M_2]$, $a \in [N_1^i]$, $b \in [N_2^j]$, $k\in [M_1]$, $l\in [M_2]$. Note that sets denoted by normal fonts consist of $n$-length sequences, while sets denoted by calligraphic fonts consist of joint distributions. The expression $TH$ refers to the word "threshold".

\begin{align}
&TH_{a,b}^{i,j}\triangleq\left\{\hspace{-4pt}\begin{array}{l}
  \vz \in \mathcal{Z}^n: \I(\vx_{a}^{i} \wedge \vy_{b}^{j} \wedge \vz | \vu) -R_1^i-R_2^j \le \eta_n \text{ or} \\
  \ \quad \I(\vx_{a}^{i} \wedge \vy_{b}^{j}, \vz | \vu) -R_1^i \le \eta_n \text{ or} \\
  \ \quad \I(\vy_{b}^{j} \wedge \vx_{a}^{i}, \vz | \vu) -R_2^j \le \eta_n.
  \end{array}\hspace{-4pt}\right\} \label{defo} \\
&V_{a,b}^{i,j} \triangleq\left\{\hspace{-4pt}\begin{array}{l}
  \vz \in \mathcal{Z}^n: \alpha(\vu, \vx_{a}^{i},\vy_{b}^{j},\vz)-R_1^i-R_2^j  \\
  \qquad\ge \alpha(\vu, \vx_{c}^{k},\vy_{d}^{l},\vz) -R_1^k-R_2^l, \text{for all} \\
  \ \quad k\in [M_1], l\in [M_2], c \in [N_1^k], d \in [N_2^l].
  \end{array}\hspace{-4pt}\right\}
\end{align}
\begin{align}
&\mathcal{TH}^{i,j}\triangleq\left\{\hspace{-3pt}
\begin{array}{l}
  V_{UXYZ} \in \mathcal{P}(\mathcal{U} \times \mathcal{X} \times \mathcal{Y} \times \mathcal{Z}) : \\
  V_{UX}=P_UP^{i}_{X|U},\,V_{UY}=P_UP^{j}_{Y|U} \\
  \I_{V}(X \wedge Y, Z | U) -R_1^i \le \eta_n \text{ or} \\
  \I_{V}(Y \wedge X, Z | U) -R_2^j \le \eta_n  \text{ or} \\
  \I_{V}(X \wedge Y \wedge Z | U) -R_1^i-R_2^j \le \eta_n.
\end{array}\hspace{-3pt}\right\} \\
&\mathcal{VX}_{k}^{i,j} \triangleq\left\{\hspace{-6pt}\begin{array}{l}
V_{UX\tilde{X}YZ} \in \mathcal{P}(\mathcal{U} \times \mathcal{X} \times \mathcal{X} \times \mathcal{Y} \times \mathcal{Z}): \\
\alpha(V_{UXYZ}) - R_1^i \le \alpha(V_{U\tilde{X}YZ})-R_1^k, \\
V_{UX}=P_UP^{i}_{X|U},\quad V_{U\tilde{X}}=P_UP^{k}_{X|U},\\
V_{UY}=P_UP^{j}_{Y|U}.
\end{array}\hspace{-6pt}\right\} \label{set3} \\
&\mathcal{VY}_{l}^{i,j} \triangleq\left\{\hspace{-6pt}\begin{array}{l}
  V_{UXY\tilde{Y}Z} \in \mathcal{P}(\mathcal{U} \times \mathcal{X} \times \mathcal{Y} \times \mathcal{Y} \times \mathcal{Z}): \\
  \alpha(V_{UXYZ}) - R_2^j \le \alpha(V_{UX\tilde{Y}Z})-R_2^l, \\
  V_{UX}=P_UP^{i}_{X|U},V_{UY}=P_UP^{j}_{Y|U},\\
  V_{U\tilde{Y}}=P_UP^{l}_{Y|U}.
  \end{array}\hspace{-6pt}\right\} \label{set4} \\
&\mathcal{VXY}_{k,l}^{i,j} \triangleq\left\{\hspace{-6pt}\begin{array}{l}
  V_{UX\tilde{X}Y\tilde{Y}Z} \in \mathcal{P}(\mathcal{U} \times \mathcal{X} \times \mathcal{X} \times \mathcal{Y} \times \mathcal{Y} \times \mathcal{Z}):  \\
  \alpha(V_{UXYZ}) - R_1^i-R_2^j \\
  \qquad\le \alpha(V_{U\tilde{X}\tilde{Y}Z})-R_1^k-R_2^l, \\
  V_{UX}=P_U P^{i}_{X|U},\quad V_{U\tilde{X}}=P_U P^{k}_{X|U},\\
  V_{UY}=P_U P^{j}_{Y|U},\quad V_{U\tilde{Y}}=P_U P^{l}_{Y|U}.
  \end{array}\hspace{-6pt}\right\}\label{set5}
\end{align}
\begin{align}
&\mathcal{TH}^{n,i,j}\triangleq \mathcal{TH}^{i,j} \cap \mathcal{P}^n (\mathcal{U} \times \mathcal{X} \times \mathcal{Y} \times \mathcal{Z}) \\
&\mathcal{VX}_{k}^{n,i,j}\triangleq \mathcal{VX}_{k}^{i,j} \cap \mathcal{P}^n (\mathcal{U} \times \mathcal{X} \times \mathcal{X} \times \mathcal{Y} \times \mathcal{Z}) \\
&\mathcal{VY}_{l}^{n,i,j}\triangleq \mathcal{VY}_{l}^{i,j} \cap \mathcal{P}^n (\mathcal{U} \times \mathcal{X} \times \mathcal{Y} \times \mathcal{Y} \times \mathcal{Z}) \\
&\mathcal{VXY}_{k,l}^{n,i,j} \triangleq \mathcal{VXY}_{k,l}^{i,j} \cap \mathcal{P}^n (\mathcal{U} \times \mathcal{X} \times \mathcal{X} \times \mathcal{Y} \times \mathcal{Y} \times \mathcal{Z})
\end{align}
\begin{align}
&\mathcal{E}TH^{i,j}\triangleq \min_{V_{UXYZ} \in \mathcal{TH}^{i,j}} \Big[ \DD(V_{Z|UXY} ||W|V_{UXY}) +\I_{V}(X \wedge Y|U) \Big] \label{ex0} \\
&\mathcal{E}X_{k}^{i,j} \triangleq \min_{V_{UX\tilde{X}YZ} \in \mathcal{VX}_{k}^{i,j}} \Big[ \DD(V_{Z|UXY} ||W|V_{UXY}) +\I_{V}(X \wedge Y|U) + |\I_{V}(\tilde{X} \wedge X,Y,Z|U)-R_1^k|^{+} \Big]  \label{ex1} \\
&\mathcal{E}Y_{l}^{i,j} \triangleq \min_{V_{UXY\tilde{Y}Z} \in \mathcal{VY}_{l}^{i,j}} \Big[ \DD(V_{Z|UXY} ||W|V_{UXY}) +\I_{V}(X \wedge Y|U) + |\I_{V}(\tilde{Y} \wedge X,Y,Z|U)-R_2^l|^{+} \Big] \label{ex2} \\
&\mathcal{E}XY_{k,l}^{i,j} \triangleq \min_{V_{UX\tilde{X}Y\tilde{Y}Z} \in \mathcal{VXY}_{k,l}^{i,j}} \Big[ \DD(V_{Z|UXY} ||W|V_{UXY})+\I_{V}(X \wedge Y|U) \notag\\ &+\Big|\I_{V}(\tilde{X}, \tilde{Y} \wedge X,Y,Z|U) +\I_{V}(\tilde{X} \wedge \tilde{Y}|U)-R_1^k-R_2^l\Big|^{+} \Big] \label{ex3}
\end{align}

\emph{Step 2: Relating error probabilities to packing functions}

For all $(i,j) \in [M_1] \times [M_2]$ we have
\begin{align}
&Err_d(i,j) \le \frac{1}{N_1^i N_2^j} \sum_{a=1}^{N_1^i} \sum_{b=1}^{N_2^j} W^{n} \left( TH_{a,b}^{i,j}  | \vx_{a}^{i},\vy_{b}^{j} \right) \label{hibatagok}\\
&\qquad+\frac{1}{N_1^i N_2^j} \sum_{a=1}^{N_1^i} \sum_{b=1}^{N_2^j} W^{n} \left( \bigcup_{\sumfrac{c=1}{c \ne a}}^{N_1^i} V_{c,b}^{i,j}  | \vx_{a}^{i},\vy_{b}^{j}\right)  \notag\\
&\qquad+\frac{1}{N_1^i N_2^j} \sum_{a=1}^{N_1^i} \sum_{b=1}^{N_2^j} W^{n} \left( \bigcup_{\sumfrac{d=1}{d \ne b}}^{N_2^j} V_{a,d}^{i,j}  | \vx_{a}^{i},\vy_{b}^{j} \right)  \notag\\
&\qquad+\frac{1}{N_1^i N_2^j} \sum_{a=1}^{N_1^i} \sum_{b=1}^{N_2^j} W^{n} \left( \bigcup_{\sumfrac{c=1}{c \ne a}}^{N_1^i}\bigcup_{\sumfrac{d=1}{d \ne b}}^{N_2^j} V_{c,d}^{i,j}  | \vx_{a}^{i},\vy_{b}^{j} \right)  \notag\\
&\qquad+\sum_{\sumfrac{k=1}{ k\ne i}}^{M_1} \frac{1}{N_1^i N_2^j} \sum_{a=1}^{N_1^i} \sum_{b=1}^{N_2^j} W^{n} \left( \bigcup_{c=1}^{N_1^k} V_{c,b}^{k,j}  | \vx_{a}^{i},\vy_{b}^{j} \right)  \notag\\
&\qquad+\sum_{\sumfrac{l=1} {l \ne j }}^{M_2} \frac{1}{N_1^i N_2^j} \sum_{a=1}^{N_1^i} \sum_{b=1}^{N_2^j} W^{n} \left( \bigcup_{d=1}^{N_2^l} V_{a,d}^{i,l}  | \vx_{a}^{i},\vy_{b}^{j} \right)  \notag\\
&\qquad+\sum_{\sumfrac{k=1}{ k\ne i}}^{M_1}\sum_{\sumfrac{l=1} {l \ne j }}^{M_2} \frac{1}{N_1^i N_2^j} \sum_{a=1}^{N_1^i} \sum_{b=1}^{N_2^j} W^{n} \left( \bigcup_{c=1}^{N_1^k}\bigcup_{d=1}^{N_2^l} V_{c,d}^{k,l}  | \vx_{a}^{i},\vy_{b}^{j} \right). \notag
\end{align}
For the sake of brevity, we introduce the following notations for the terms of the right-hand side of equation (\ref{hibatagok}):
\begin{align}
&Err_d(i,j) \le th^{i,j} + errorX^{i,j}_{i} + errorY^{i,j}_{j} +errorXY_{i,j}^{i,j} \notag\\
&\qquad + \sum_{\sumfrac{k=1}{ k\ne i}}^{M_1} errorX_{k}^{i,j} +\sum_{\sumfrac{l=1} {l \ne j }}^{M_2} errorY_{l}^{i,j} + \sum_{\sumfrac{k=1}{ k\ne i}}^{M_1}\sum_{\sumfrac{l=1} {l \ne j }}^{M_2} errorXY_{k,l}^{i,j} . \label{hibatagok2}
\end{align}

In words, $th^{i,j}$ is the average probability of the event that the sent codewords do not fulfill the threshold criterion (\ref{crit1})-(\ref{crit3}) when codebook pair $(i,j)$ was chosen by the senders. The other terms correspond to the average probability that the sent codewords do not uniquely maximize (\ref{maximum}).

We claim that the terms of (\ref{hibatagok2}) can be bounded via packing functions of Lemma \ref{Rc-Packing-lemma} in this way:
\begin{align}
&th^{i,j} \le \sum_{\sumfrac{V_{UXYZ} \in}{\mathcal{TH}^{n,i,j}}} \big( 2^{-n[\DD(V_{Z|UXY}||W|V_{UXY})]}  \frac{1}{N_1^i N_2^j} K^{i,j}(V_{UXY}) \big) \label{hibath} \\
&errorX_{k}^{i,j} \le \sum_{\sumfrac{V_{UX\tilde{X}YZ} \in}{\mathcal{VX}_{k}^{n,i,j}}} \big( 2^{-n[\DD(V_{Z|UXY}||W|V_{UXY})]}  \frac{1}{N_1^i N_2^j}   min \{2^{-n \I_{V}(\tilde{X}\wedge Z|XYU)}K_{k}^{i,j}(V_{UX\tilde{X}Y}),K^{i,j}(V_{UXY}) \} \big) \label{hibaX} \\
&errorY_{l}^{i,j} \le \sum_{\sumfrac{V_{UXY\tilde{Y}Z} \in}{\mathcal{VY}_{l}^{n,i,j}}} \big( 2^{-n[\DD(V_{Z|UXY}||W|V_{UXY})]}  \frac{1}{N_1^i N_2^j}  min \{2^{-n \I_{V}(\tilde{Y}\wedge Z|XYU)}K_{l}^{i,j}(V_{UXY\tilde{Y}}),K^{i,j}(V_{UXY}) \}\big) \label{hibaY} \\
&errorXY_{k,l}^{i,j} \le \sum_{\sumfrac{V_{UX\tilde{X}Y\tilde{Y}Z} \in}{\mathcal{VXY}_{k,l}^{n,i,j}}} \big( 2^{-n[\DD(V_{Z|UXY}||W|V_{UXY})]} \frac{1}{N_1^i N_2^j}  min \{2^{-n \I_{V}(\tilde{X}\tilde{Y}\wedge Z|XYU)}K_{k,l}^{i,j}(V_{UX\tilde{X}Y\tilde{Y}}),K^{i,j}(V_{UXY}) \}\big) \label{hibaXY}
\end{align}
First let us prove inequality (\ref{hibath}).
\begin{align}
&th^{i,j} = \frac{1}{N_1^i N_2^j} \sum_{a=1}^{N_1^i} \sum_{b=1}^{N_2^j} W^{n} \left( TH_{a,b}^{i,j}  | \vx_{a}^{i},\vy_{b}^{j} \right) \label{ideiskell1}\\
&\le \sum_{\sumfrac{V_{UXYZ} \in}{\mathcal{TH}^{n,i,j}}} \sum_{a=1}^{N_1^i} \sum_{b=1}^{N_2^j} \Big( 2^{-n[\DD(V_{Z|UXY}||W|V_{UXY}) + \HH_{V}(Z|UXY)]} \frac{1}{N_1^i N_2^j} \big| \vz \in \mathcal{Z}^n: P_{(\vu, \vx_{a}^i, \vy_{b}^j, \vz)}= V \big| \Big), \label{ideiskell2}
\end{align}
In (\ref{ideiskell2}) we have used (\ref{basicfact4}). Using (\ref{basicfact3}) we obtain
\begin{align}
&\big| \vz \in \mathcal{Z}^n: P_{(\vu, \vx_{a}^i, \vy_{b}^j, \vz)}= V \big| \le \mathds 1_{T^{n}_{V_{UXY}}}(\vu,\vx_{a}^{i},\vy_{b}^{j}) \cdot 2^{n[ \HH_{V}(Z|UXY)]}. \label{modszeris1}
\end{align}
Substituting inequality (\ref{modszeris1}) in (\ref{ideiskell1}) and (\ref{ideiskell2}) and taking into account the definition of packing function $K^{i,j}[V_{UXY}]$ from Lemma \ref{Rc-Packing-lemma},  the bound (\ref{hibath}) follows.

Let us prove now inequality (\ref{hibaXY}). The bounds (\ref{hibaX}) and (\ref{hibaY}) can be proved analogously.
\begin{align}
&errorXY_{k,l}^{i,j} = \frac{1}{N_1^i N_2^j} \sum_{a=1}^{N_1^i} \sum_{b=1}^{N_2^j} W^{n} \left( \bigcup_{\sumfrac{c=1}{c \ne a \textnormal{ if } i = k}}^{N_1^k}\bigcup_{\sumfrac{d=1}{d \ne b \textnormal{ if } j = l}}^{N_2^l} V_{c,d}^{k,l}  | \vx_{a}^{i},\vy_{b}^{j} \right) \label{idekell1} \\
&\le \sum_{\sumfrac{V_{UX\tilde{X}Y\tilde{Y}Z} \in}{\mathcal{VXY}_{k,l}^{n,i,j}}} \sum_{a=1}^{N_1^i} \sum_{b=1}^{N_2^j} 2^{-n[\DD(V_{Z|UXY}||W|V_{UXY}) + \HH_{V}(Z|UXY)]} \frac{1}{N_1^i N_2^j} \big| \vz \in \mathcal{Z}^n: \text{for some } (c,d) \text{ } P_{(\vu, \vx_{a}^i,\vx_{c}^k, \vy_{b}^j, \vy_{d}^l, \vz)}= V \big|. \label{idekell3}
\end{align}
In (\ref{idekell3}) we have used (\ref{basicfact4}). Using (\ref{basicfact3}) we obtain two bounds
\begin{align}
&\big| \vz \in \mathcal{Z}^n: \text{for some } (c,d) \text{ } P_{(\vu, \vx_{a}^i,\vx_{c}^k, \vy_{b}^j, \vy_{d}^l, \vz)}= V \big| \le \mathds 1_{T^{n}_{V_{UXY}}}(\vu,\vx_{a}^{i},\vy_{b}^{j}) \cdot 2^{n[ \HH_{V}(Z|UXY)]}, \label{modszer1}
\end{align}
and
\begin{align}
&\big| \vz \in \mathcal{Z}^n: \text{for some } (c,d) \text{ } P_{(\vu, \vx_{a}^i,\vx_{c}^k, \vy_{b}^j, \vy_{d}^l, \vz)}= V \big| \notag\\
&\le \sum^{N_1^k}_{\sumfrac{c=1}{c \ne a \textnormal{ if } i = k}}\sum^{N_2^l}_{\sumfrac{d=1}{d \ne b \textnormal{ if } j = l}}\mathds 1_{T^{n}_{V_{UX\tilde{X}Y\tilde{Y}}}}(\vu,\vx_{a}^{i},\vx_{c}^{k}, \vy_{b}^{j}, \vy_{d}^{l}) \cdot 2^{n[ \HH_{V}(Z|UX\tilde{X}Y\tilde{Y})]}. \label{modszer2}
\end{align}
Substituting inequalities (\ref{modszer1}), (\ref{modszer2}) in (\ref{idekell1}) and (\ref{idekell3}) and taking into account the definition of packing functions from Lemma \ref{Rc-Packing-lemma},  the bound (\ref{hibaXY}) follows.

\emph{Step 3: Error exponents of type \cite{Nazari}}

The following consequences of Lemma \ref{multiinfo} are used in this step:
\begin{align}
&\I_{V}(X \wedge \tilde{X} \wedge Y|U)=\I_{V}(\tilde{X} \wedge X,Y|U)+ \I_{V}(X \wedge Y|U), \label{multi1} \\
&\I_{V}(X \wedge Y \wedge \tilde{Y}|U)=\I_{V}(\tilde{Y} \wedge X,Y|U)+ \I_{V}(X \wedge Y|U), \label{multi2} \\
&\I_{V}(X \wedge \tilde{X} \wedge Y \wedge \tilde{Y}|U)=\I_{V}(\tilde{X}, \tilde{Y} \wedge X,Y|U) + \I_{V}(\tilde{X} \wedge \tilde{Y}|U)+ \I_{V}(X \wedge Y|U). \label{multi3}
\end{align}
Substitute $N_1^i = \left\lfloor 2^{nR_1^i} \right\rfloor$ and $N_2^j = \left\lfloor 2^{nR_2^j} \right\rfloor$ in inequalities (\ref{hibath}), (\ref{hibaX}), (\ref{hibaY}) and (\ref{hibaXY}), use Lemma \ref{Rc-Packing-lemma} to upper-bound the packing functions. Then use (\ref{basicfact1}) and the above identities (\ref{multi1})-(\ref{multi3}), take into account the uniform continuity (the alphabets are finite) of the present information measures (in order to exchange $\mathcal{TH}^{n,i,j}$ to $\mathcal{TH}^{i,j}$ and $\mathcal{VX}_{k}^{n,i,j}$ to $\mathcal{VX}_{k}^{i,j}$ etc.) and draw the following conclusions:
\begin{align}
&th^{i,j} \le 2^{-n(\mathcal{E}TH^{i,j} - \delta^{''}_{n})}, \label{Nazarifele1} \\
&errorX_{k}^{i,j} \le 2^{-n(\mathcal{E}X_{k}^{i,j} - \delta^{''}_{n})}, \text{ } errorY_{l}^{i,j} \le 2^{-n(\mathcal{E}Y_{l}^{i,j} - \delta^{''}_{n})}, \label{Nazarifele2} \\
&errorXY_{k,l}^{i,j} \le 2^{-n(\mathcal{E}XY_{k,l}^{i,j} - \delta^{''}_{n})} \label{Nazarifele3}
\end{align}
for some sequence $\delta^{''}_{n}(|\mathcal{U}|,|\mathcal{X}|,|\mathcal{Y}|,|\mathcal{Z}|)$ $\rightarrow 0$.

\emph{Step 4: Final calculations}

The following inequalities are consequences of the chain rule for mutual information and Lemma \ref{multiinfo}:
\begin{align}
&\I_{V}(\tilde{X} \wedge X,Y,Z|U) \ge \I_{V}(\tilde{X} \wedge Y,Z|U), \label{chain1} \\
&\I_{V}(\tilde{Y} \wedge X,Y,Z|U) \ge \I_{V}(\tilde{Y} \wedge X,Z|U), \label{chain2} \\
&\I_{V}(\tilde{X},\tilde{Y} \wedge X,Y,Z|U) + \I_{V}(\tilde{X} \wedge \tilde{Y}|U) \ge \I_{V}(\tilde{X},\tilde{Y} \wedge Z|U) + \I_{V}(\tilde{X} \wedge \tilde{Y}|U) \label{chain3}= \I_{V}(\tilde{X} \wedge \tilde{Y} \wedge Z|U).
\end{align}

Via inequalities (\ref{chain1}), (\ref{chain2}) and (\ref{chain3}) the exponents $\mathcal{E}X_{k}^{i,j}$, $\mathcal{E}Y_{l}^{i,j}$, $\mathcal{E}XY_{k,l}^{i,j}$ can be bounded from below in the following way:
\begin{align}
&\mathcal{E}X_{k}^{i,j} \ge \min_{V_{UX\tilde{X}YZ} \in \mathcal{VX}_{k}^{i,j}} \Big[ \DD(V_{Z|UXY} ||W|V_{UXY}) +\I_{V}(X \wedge Y|U) + |\I_{V}(\tilde{X} \wedge Y,Z|U)-R_1^k|^{+} \Big], \label{felighl1} \\
&\mathcal{E}Y_{l}^{i,j} \ge  \min_{V_{UXY\tilde{Y}Z} \in \mathcal{VY}_{l}^{i,j}} \Big[  \DD(V_{Z|UXY} ||W|V_{UXY}) +\I_{V}(X \wedge Y|U) + |\I_{V}(\tilde{Y} \wedge X,Z|U)-R_2^l|^{+} \Big],  \label{felighl2} \\
&\mathcal{E}XY_{k,l}^{i,j} \ge \min_{V_{UX\tilde{X}Y\tilde{Y}Z} \in \mathcal{VXY}_{k,l}^{i,j}} \Big[  \DD(V_{Z|UXY} ||W|V_{UXY}) + \I_{V}(X \wedge Y|U) +|\I_{V}(\tilde{X} \wedge \tilde{Y} \wedge Z|U) -R_1^k-R_2^l|^{+} \Big] . \label{felighl3}
\end{align}

If we examine definitions (\ref{set3})-(\ref{set5}) and take into account Lemma \ref{multiinfo} we can draw the following two conclusions. If $V_{UX\tilde{X}YZ} \in \mathcal{VX}_{k}^{i,j}$ then $I_{V}(\tilde{X} \wedge Y,Z|U)-R_1^k$ is not smaller than $I_{V}(X \wedge Y,Z|U)-R_1^i$. If $V_{UXY\tilde{Y}Z} \in \mathcal{VY}_{l}^{i,j}$ then $I_{V}(\tilde{Y} \wedge X,Z|U)-R_2^l$ is not smaller than $I_{V}(Y \wedge X,Z|U)-R_2^j$. Using these conclusions to lower-bound the right hand sides of (\ref{felighl1})-(\ref{felighl3}) leads to the following inequalities
\begin{align}
&\mathcal{E}X_{k}^{i,j} \ge \mathcal{E}X_{LH}^{i,j},\quad \mathcal{E}Y_{k,l}^{i,j} \ge \mathcal{E}Y_{LH}^{i,j}, \quad \mathcal{E}XY_{k,l}^{i,j} \ge \mathcal{E}XY_{LH}^{i,j}, \label{hugesliu}
\end{align}
where for any $\beta \in \{X,Y,XY \}$ the expression $\mathcal{E}\beta_{LH}^{i,j}$ is equal to $\mathcal{E}\beta_{LH}(R_1^i,R_2^j,W,P_{U},P^i_{X|U},P^j_{Y|U})$.

Now we examine the relationship between exponent $\mathcal{E}TH^{i,j}$ and exponents $\mathcal{E}X_{LH}^{i,j}$,  $\mathcal{E}Y_{LH}^{i,j}$, $\mathcal{E}XY_{LH}^{i,j}$. For the sets on which the minima are taken we have $\mathcal{TH}^{i,j} \subseteq \mathcal{V}^{i,j}_{LH}=\mathcal{V}_{LH}(P_{U}, P^i_{X|U},P^j_{Y|U})$, see (\ref{hl1}). Furthermore, for any $V_{UXYZ} \in \mathcal{TH}^{i,j}$ there exist $\beta \in \{X,Y,XY \}$ such that the difference between the optimized functions in the definition of $\mathcal{E}TH^{i,j}$ and $\mathcal{E}\beta_{LH}^{i,j}$ evaluated in $V_{UXYZ}$ is at least $\eta_n$. These considerations lead to the inequality
\begin{equation}
\min(\mathcal{E}X_{LH}^{i,j}, \mathcal{E}Y_{LH}^{i,j},\mathcal{E}XY_{LH}^{i,j}) \le \mathcal{E}TH^{i,j} + \eta_n. \label{thnemszolbele}
\end{equation}
As $M_1$ and $M_2$ grow subexponentially in $n$, part (i) of Theorem \ref{collision} is proved via inequalities  (\ref{hibatagok2}), (\ref{Nazarifele1})-(\ref{Nazarifele3}), (\ref{hugesliu}), (\ref{thnemszolbele}). Note that in this part $\eta_n(|\mathcal{U}|,|\mathcal{X}|,|\mathcal{Y}|,|\mathcal{Z}|,M_1,M_2)$ can be arbitrary positive sequence which goes to $0$ as $n \rightarrow \infty$. However, it will turn out from the proof of part (ii) that the sequence $\eta_n$ has to converge to $0$ sufficiently slowly.

\subsection*{Proof of part (ii) of Theorem \ref{collision}}

\emph{Step 1: Some definitions}

The following definitions are meant for all possible combinations of indices $i\in [M_1]$, $j\in [M_2]$, $k\in [M_1]$, $l\in [M_2]$. The symbol $C$ refers to "complement in some sense".

\begin{align}
&\mathcal{CTH}^{i,j} \triangleq\left\{\hspace{-6pt}\begin{array}{l}
V_{UXYZ} \in \mathcal{P}(\mathcal{U} \times \mathcal{X} \times \mathcal{Y} \times \mathcal{Z}): \\
V_{UX}=P_UP^{i}_{X|U},\quad V_{UY}=P_UP^{j}_{Y|U},\\
\I_{V}(X \wedge Y, Z | U) -R_1^i > \eta_n, \\
  \I_{V}(Y \wedge X, Z | U) -R_2^j > \eta_n, \\
  \I_{V}(X \wedge Y \wedge Z | U) -R_1^i-R_2^j > \eta_n.
\end{array}\hspace{-6pt}\right\} \label{setmasik0} \\
&\mathcal{CTHX}_{k}^{i,j} \triangleq\left\{\hspace{-6pt}\begin{array}{l}
V_{UX\tilde{X}YZ} \in \mathcal{P}(\mathcal{U} \times \mathcal{X} \times \mathcal{X} \times \mathcal{Y} \times \mathcal{Z}): \\
V_{UX}=P_UP^{i}_{X|U},\quad V_{U\tilde{X}}=P_UP^{k}_{X|U},\\
V_{UY}=P_UP^{j}_{Y|U}, \\
\I_{V}(\tilde{X} \wedge Y, Z | U) -R_1^k > \eta_n, \\
  \I_{V}(Y \wedge \tilde{X}, Z | U) -R_2^j > \eta_n, \\
  \I_{V}(\tilde{X} \wedge Y \wedge Z | U) -R_1^k-R_2^j > \eta_n.
\end{array}\hspace{-6pt}\right\} \label{setmasik1}
\end{align}
\begin{align}
&\mathcal{CTHY}_{l}^{i,j} \triangleq\left\{\hspace{-6pt}\begin{array}{l}
  V_{UXY\tilde{Y}Z} \in \mathcal{P}(\mathcal{U} \times \mathcal{X} \times \mathcal{Y} \times \mathcal{Y} \times \mathcal{Z}): \\
  V_{UX}=P_UP^{i}_{X|U},V_{UY}=P_UP^{j}_{Y|U},\\
  V_{U\tilde{Y}}=P_UP^{l}_{Y|U}, \\
  \I_{V}(X \wedge \tilde{Y}, Z | U) -R_1^i > \eta_n, \\
  \I_{V}(\tilde{Y} \wedge X, Z | U) -R_2^l > \eta_n, \\
  \I_{V}(X \wedge \tilde{Y} \wedge Z | U) -R_1^i-R_2^l > \eta_n.
  \end{array}\hspace{-6pt}\right\} \label{setmasik2} \\
&\mathcal{CTHXY}_{k,l}^{i,j} \triangleq\left\{\hspace{-6pt}\begin{array}{l}
  V_{UX\tilde{X}Y\tilde{Y}Z} \in \\
  \mathcal{P}(\mathcal{U} \times \mathcal{X} \times \mathcal{X} \times \mathcal{Y} \times \mathcal{Y} \times \mathcal{Z}):  \\
  V_{UX}=P_U P^{i}_{X|U},\quad V_{U\tilde{X}}=P_U P^{k}_{X|U},\\
  V_{UY}=P_U P^{j}_{Y|U},\quad V_{U\tilde{Y}}=P_U P^{l}_{Y|U}, \\
  \I_{V}(\tilde{X} \wedge \tilde{Y}, Z | U) -R_1^k > \eta_n, \\
  \I_{V}(\tilde{Y} \wedge \tilde{X}, Z | U) -R_2^l > \eta_n, \\
  \I_{V}(\tilde{X} \wedge \tilde{Y} \wedge Z | U) -R_1^k-R_2^l > \eta_n.
   \end{array}\hspace{-6pt}\right\}\label{setmasik3}
\end{align}
\begin{align}
&\mathcal{E}CTH^{i,j}\triangleq \inf_{V_{UXYZ} \in \mathcal{CTH}^{i,j}} \Big[  \DD(V_{Z|UXY} ||W|V_{UXY}) +\I_{V}(X \wedge Y|U) \Big] \label{ex4} \\
&\mathcal{E}CTHX^{i,j}_{k} \triangleq \inf_{V_{UX\tilde{X}YZ} \in \mathcal{CTHX}_{k}^{i,j}} \Big[  \DD(V_{Z|UXY} ||W|V_{UXY}) +\I_{V}(X \wedge Y|U) + |\I_{V}(\tilde{X} \wedge X,Y,Z|U)-R_1^k|^{+} \Big]  \label{ex5} \\
&\mathcal{E}CTHY^{i,j}_{l} \triangleq \inf_{V_{UXY\tilde{Y}Z} \in \mathcal{CTHY}_{l}^{i,j}} \Big[  \DD(V_{Z|UXY} ||W|V_{UXY})+\I_{V}(X \wedge Y|U) + |\I_{V}(\tilde{Y} \wedge X,Y,Z|U)-R_2^l|^{+} \Big]   \label{ex6} \\
&\mathcal{E}CTHXY^{i,j}_{k,l} \triangleq \inf_{V_{UX\tilde{X}Y\tilde{Y}Z} \in \mathcal{CTHXY}_{k,l}^{i,j}} \Big[  \DD(V_{Z|UXY} ||W|V_{UXY}) +\I_{V}(X \wedge Y|U)\notag\\ &+\Big|\I_{V}(\tilde{X}, \tilde{Y} \wedge X,Y,Z|U)+\I_{V}(\tilde{X} \wedge \tilde{Y}|U)-R_1^k-R_2^l\Big|^{+} \Big]  \label{ex7}
\end{align}

\emph{Step 2: Calculations}
\begin{align}
&Err_c(i,j) \le \frac{1}{N_1^i N_2^j} \sum_{a=1}^{N_1^i} \sum_{b=1}^{N_2^j} W^{n} \Big(\mathbf{z}: \text{for some} k\in [M_1], l\in [M_2], c \in [N_1^k], d \in [N_2^l], \text{ } \mathbf{z} \notin TH_{c,d}^{k,l}  | \vx_{a}^{i},\vy_{b}^{j} \Big). \label{ec}
\end{align}
Using union bound, it is possible to expand (\ref{ec}) similarly as $Err_d(i,j)$ is expanded in (\ref{hibatagok}). The differences are that instead of $TH_{a,b}^{i,j}$ its complement should be taken and  instead of the set $V_{c,d}^{k,l}$ the complement of the set $TH_{c,d}^{k,l}$ should be taken for all $(k,c,l,d) \ne (i,a,j,b)$.

Replicating the proof of part (i) leads to the following statements. For some sequence $\delta^{'''}_{n}(|\mathcal{U}|,|\mathcal{X}|,|\mathcal{Y}|,|\mathcal{Z}|)$ $\rightarrow 0$, the first term in the upper bound of $Err_c(i,j)$ (where the complement of $TH_{a,b}^{i,j}$ is present) can be upper-bounded by
\begin{equation} \label{exponensekcol0}
2^{-n(\mathcal{E}CTH^{i,j} - \delta^{'''}_{n})},
\end{equation}
while the sum of the other terms can be upper-bounded by the following expression:
\begin{align}
&\sum_{k=1}^{M_1} 2^{-n(\mathcal{E}CTHX^{i,j}_{k} - \delta^{'''}_{n})} + \sum_{l=1}^{M_2} 2^{-n(\mathcal{E}CTHY^{i,j}_{l} - \delta^{'''}_{n})} +\sum_{k=1}^{M_1}\sum_{l=1}^{M_2} 2^{-n(\mathcal{E}CTHXY^{i,j}_{k,l} - \delta^{'''}_{n})}. \label{exponensekcol}
\end{align}
It remains to show that (\ref{exponensekcol0}) and (\ref{exponensekcol}) approach $0$ if $\eta_{n} >0$ goes to $0$ sufficiently slowly.

Using the definitions of the sets (\ref{setmasik1}), (\ref{setmasik2}) and (\ref{setmasik3}), in expressions (\ref{ex5}), (\ref{ex6}), and (\ref{ex7}) the positive part signs $|\dots|^{+}$ can be lower-bounded by $\eta_n$. As $M_1$, $M_2$ grow at most subexponentially in $n$ we are done with expression (\ref{exponensekcol}).

To prove rigorously the claim about (\ref{exponensekcol0}) we need the following argument. In part (ii) of Theorem \ref{collision} it is assumed that  $(R_1^{i},R_2^{j})$ is not in the interior of $C[W,P_{U},P^{i}_{X|U},P^{j}_{Y|U}]$. Hence, the distribution $P_{U}P^{i}_{X|U}P^{j}_{Y|U}W$ is in $\mathcal{TH}^{i,j}$. Actually for each $V_{UXYZ} \in \mathcal{CTH}^{i,j}$, one of the following inequalities holds
\begin{align}
& \I_{V}(X \wedge Y,Z|U) - \I_{P_{U}P^{i}_{X|U}P^{j}_{Y|U}W}(X \wedge Y,Z|U) > \eta_n, \\
&  \I_{V}(Y \wedge X,Z|U) - \I_{P_{U}P^{i}_{X|U}P^{j}_{Y|U}W}(Y \wedge X,Z|U) > \eta_n, \\
&  \I_{V}(X \wedge Y \wedge Z|U) - \I_{P_{U}P^{i}_{X|U}P^{j}_{Y|U}W}(X \wedge Y \wedge Z|U) > \eta_n.
\end{align}
Using uniform continuity of mutual information, it follows that there exists some $\gamma(\eta_n,|\mathcal{U}|,|\mathcal{X}|,|\mathcal{Y}|,|\mathcal{Z}|)>0$ such that if $V_{UXYZ} \in \mathcal{CTH}^{i,j}$ then its variational distance from $P_{U}P^{i}_{X|U}P^{j}_{Y|U}W$ is at least $\gamma$. Note that if $V_{UXYZ} \in \mathcal{CTH}^{i,j}$ then $V_{U}=P_{U}$ by definition, hence $||V_{UXYZ}-P_{U}P^{i}_{X|U}P^{j}_{Y|U}W||$ is equal to
\begin{align}
& \sum_{u \in \mathcal{U}} P_{U}(u) \sum_{(x,y,z) \in \mathcal{X} \times \mathcal{Y} \times \mathcal{Z}}   |V_{XYZ|U=u} (x,y,z)-  P^{i}_{X|U=u}P^{j}_{Y|U=u}W (x,y,z)|.  \label{TVatir}
\end{align}
It follows that for each $V_{UXYZ} \in \mathcal{CTH}^{i,j}$ there exists $u \in \mathcal{U}$ such that the variational distance of $V_{XYZ|U=u}$ and $P^{i}_{X|U=u}P^{j}_{Y|U=u}W$ is at least $\frac{\gamma}{|\mathcal{U}|P_{U}(u)}$. Note also that in (\ref{ex4}), $\DD(V_{Z|UXY} ||W|V_{UXY})+\I_{V}(X \wedge Y|U)$ is equal to $\DD(V_{XYZ|U} ||P^{i}_{X|U}P^{j}_{Y|U}W|P_U )$. Hence, taking into account the Pinsker inequality and (\ref{Kullback}) the claim about (\ref{exponensekcol0}) follows.

Altogether, part (ii) of Theorem \ref{collision} is proved.

\section{Nonexistence of collision error probability exponent in the proof of Theorem \ref{collision}} \label{elesseg}
The following proposition shows that (\ref{exponensekcol}) does not go to zero exponentially fast, under very reasonable assumptions on the codebook library pair.

\newpage

\begin{all} \label{elessegprop}
Let us assume that a positive real number $\eta_n$ and a constant composition codebook-library pair as in Definition \ref{constantcomposition} is given such that for some $i \in [M_1]$, $j \in [M_2]$, $k \in [M_1]$ the conditional distributions $P^{i}_{X|U}$ and $P^{k}_{X|U}$ are equal, the rate pair $(R_1^i,R_2^j)$ is not in the interior of $C[W,P_{U},P^{i}_{X|U},P^{j}_{Y|U}]$, and
\begin{align}
&\I_{P_{U}P^{k}_{X|U}P^{j}_{Y|U}W} (X \wedge Y,Z|U) -R_1^k > \eta_n \label{mestersegesfeltetel0}, \\
&\I_{P_{U}P^{k}_{X|U}P^{j}_{Y|U}W} (Y \wedge Z|U) -R_1^k-R_2^j >\eta_n \label{mestersegesfeltetel}.
\end{align}
Then $\mathcal{E}CTHX^{i,j}_{k}$ defined by (\ref{setmasik1}) and (\ref{ex5}) is equal to $\eta_n$.
\end{all}
\begin{Rem}
Inequalities (\ref{mestersegesfeltetel0}) and (\ref{mestersegesfeltetel}) imply that the rate pair $(R_1^k,R_2^j)$ is in the interior of $C[W,P_{U},P^{k}_{X|U},P^{j}_{Y|U}]$.
\end{Rem}
\begin{IEEEproof}
Let the joint distribution of $(U,X,Y,Z)$ be equal to $P_{U}P^{i}_{X|U}P^{j}_{Y|U}W$. Let $\hat{X}^{*}$ be independent of $X$ given $UYZ$ with conditional distribution on $UYZ$ equal to the conditional distribution of $X$ on $UYZ$. Furthermore, let $\hat{X}^{**}$ be independent of $(X,Y,Z)$ given $U$ with conditional distribution on $U$ equal to the conditional distribution of $X$ on $U$. Let us denote the joint distribution of $(U,X,\hat{X}^{*},Y,Z)$ and $(U,X,\hat{X}^{**},Y,Z)$ by $V^{*}$ and $V^{**}$ respectively. For all $\varepsilon \in [0,1]$ define joint distribution $V^{\varepsilon}$ by $(1-\varepsilon)V^{*}+ \varepsilon V^{**}$.

For $\varepsilon$ equals to $0$ or $1$, and hence for all $\varepsilon \in [0,1]$ it follows that $\DD(V^{\varepsilon}_{Z|UXY} ||W|V^{\varepsilon}_{UXY})+ \I_{V^{\varepsilon}}(X \wedge Y|U)$ and $\I_{V^{\varepsilon}}(\tilde{X} \wedge X|U,Y,Z)$ are equal to $0$. Keeping in mind (\ref{ex5}), define $r(\varepsilon)$ by
\begin{equation}
\I_{V^{\varepsilon}}(\tilde{X} \wedge X,Y,Z|U)-R_1^k=\I_{V^{\varepsilon}}(\tilde{X} \wedge Y,Z|U)-R_1^k.
\end{equation}
Then $r(0) > \eta_n$ by (\ref{mestersegesfeltetel0}), and $r(1)$ equals $-R_1^k$. Using continuity we can find $\varepsilon_0$ with difference $r(\varepsilon_0) - \eta_n >0$ arbitrary small. Note that by the chain rule for mutual information and Lemma \ref{multiinfo} both $\I_{V^{\varepsilon_0}}(Y \wedge \tilde{X},Z|U)$ and  $\I_{V^{\varepsilon_0}}(\tilde{X} \wedge Y \wedge Z|U)$ are not less than $\I_{V^{\varepsilon_0}}(Y \wedge Z|U)$. Hence, using condition (\ref{mestersegesfeltetel}) it follows that $V^{\varepsilon_0}$ is in set (\ref{setmasik1}). Taking into account (\ref{ex5}), this proposition is proved. \end{IEEEproof}

\section*{Acknowledgment}
The preparation of this article would not have been possible without the
support of Prof. Imre Csisz{\'a}r. We would like to thank him for his help and advice within this subject area. We are also grateful to Himanshu Tyagi for useful discussions on the topic of this paper.


\begin{thebibliography}{70}
\bibitem{Borade}
\newblock S. Borade, B. Nakiboglu, L. Zheng
\newblock  ``Unequal Error Protection: An Information-Theoretic Perspective,''
\newblock \emph{IEEE Trans. Inf. Theory}, vol. 55, pp. 5511-5539, Dec. 2009.
\bibitem{campo}
\newblock A. T. Campo, G. Vazquez-Vilar, A. G. F\`abregas, T. Koch, A. Martinez
\newblock ``Joint Source-Channel Coding Revisited: Random-Coding Bounds and Error Exponents,''
\newblock \emph{arXiv:}1303.6249
\bibitem{remi}
\newblock R. A. Chou, M. R. Bloch
\newblock ``Data Compression with Nearly Uniform Output,''
\newblock \emph{in Information Theory Proceedings (ISIT), 2013 IEEE
International Symposium on}, pp. 1979-1983.
\bibitem{Csiszar}
\newblock I. Csisz{\'a}r,
\newblock ``Joint Source-Channel Error Exponent,''
\newblock \emph{Prob. Contr. \& Info. Theory}, vol. 9, no. 5, pp.315--323, 1980.
\bibitem{alpha}
\newblock I. Csisz\'ar, J. K\"orner
\newblock  ``Graph decomposition: A new key to coding theorems,''
\newblock \emph{IEEE Trans. Inf. Theory}, vol. 1, pp. 5-12, Jan. 1981.
\bibitem{Csiszar2}
\newblock I. Csisz{\'a}r, J. K{\"o}rner,
\newblock \textit{Information theory, Coding theorems for Discrete Memoryless Systems, $2^{nd}$ edition}, Cambridge University Press, 2011.
\bibitem{mi}
\newblock L. Farkas, T. Kói,
\newblock ``Random Access and Source-Channel Coding Error Exponents for Multiple Access Channels,''
\newblock \emph{in Information Theory Proceedings (ISIT), 2013 IEEE
International Symposium on}, pp. 374-378.
\bibitem{Hughes}
Y. Liu, B.L. Hughes
\newblock ``A new universal random coding bound for the multiple-access channel,''
\newblock \emph{IEEE Trans. Inf. Theory}, vol. 42, pp. 376-386 Mar. 1996.
\bibitem{Luo}
\newblock J. Luo
\newblock ``A Generalized Channel Coding Model for Random Multiple Access Communication,''
\newblock \emph{arXiv:}1306.1632
\bibitem{Ephremides}
\newblock J. Luo, A. Ephremides
\newblock ``A New Approach to Random Access: Reliable Communication and Reliable Collision Detection,''
\newblock \emph{IEEE Trans. Inf. Theory}, vol. 58, pp. 989-1002 Feb. 2012.
\bibitem{Nazari}
\newblock A. Nazari, A. Anastasopoulos, S. S. Pradhan
\newblock ``Error exponent for multiple-access channels: lower bounds,''
\newblock \emph{IEEE Trans. Inf. Theory} to appear.
\bibitem{yanina}
\newblock Y. Y. Shkel, V. Y. F. Tan, S. C. Draper
\newblock ``Converse Bounds for Assorted Codes in the Finite Blocklength Regime,''
\newblock \emph{in Information Theory Proceedings (ISIT), 2013 IEEE
International Symposium on}, pp. 1720-1724.
\bibitem{Yeung}
\newblock R.W. Yeung,
\newblock \textit{ A First Course in Information Theory},  Kluwer: New York, 2002.
\bibitem{dispersioncikk}
\newblock D. Wang, A. Ingber, Y. Kochman
\newblock ``The dispersion of joint source channel coding,''
\newblock \emph{in Communication, Control, and Computing (Allerton), 2011 49th Annual Allerton Conference on}, pp. 180-187.
\bibitem{Jie-Luo}
\newblock Z. Wang, J. Luo,
\newblock ``Error Performance of Channel Coding in Random-Access Communication,''
\newblock \emph{IEEE Trans. Inf. Theory} vol. 58, pp. 3961-3974 Jun. 2012.
\bibitem{watanabe}
\newblock S. Watanabe,
\newblock ``Information-theoretical analysis of multivariate correlation,''
\newblock \emph{IBM J. Res. Develop.}, vol. 4, pp. 66-82, 1960.
\bibitem{Alajaji}
\newblock Y. Zhong, F. Alajaji, L. L. Campbell
\newblock  ``Error Exponents for Asymmetric Two-User Discrete Memoryless Source-Channel Coding Systems,''
\newblock \emph{IEEE Trans. Inf. Theory}, vol. 55, pp. 1487-1518, April 2009.
\end{thebibliography}
\end{document}